# Microstructural and Micromechanical Evolution of Olivine Aggregates During Transient Creep

Harison S. Wiesman[1]*, Thomas Breithaupt[1], David Wallis[1], Lars N. Hansen[2]

*Corresponding Author

[1] Department of Earth Sciences, University of Cambridge, Cambridge, UK

[2] Department of Earth and Envirnomental Sciences, University of Minnesota – Twin Cities, Minneapolis, MN, USA

**Key Points**

1. Dislocation density increases with strain during transient creep and is correlated with increases in intragranular stress heterogeneity.
2. Maps of intragranular stress heterogeneity reveal that long-range dislocation interactions control the stress evolution during the transient.
3. Recent models could not reproduce the evolution of stress and dislocation density during transient creep, demanding these models be revisited.

**Abstract**

To examine the microstructural evolution that occurs during transient creep, we deformed olivine aggregates to different strains that spanned the initial transient deformation. Two sets of samples with different initial grain sizes of 5 µm and 20 µm were deformed in torsion at $T = 1523$ K, $P = 300$ MPa, and a constant shear strain rate of $1.5 \times 10^{-4}$ s$^{-1}$. Both sets of samples experienced strain hardening during deformation. We characterized the microstructures at the end of each experiment using high-angular resolution electron backscatter diffraction (HR-EBSD) and dislocation decoration. In the coarse-grained samples, dislocation density increased from $1.5 \times 10^{11}$ m$^{-2}$ to $3.6 \times 10^{12}$ m$^{-2}$ with strain. Although the same final dislocation density was reached in the fine-grained samples, it did not vary significantly at small strains, potentially due to concurrent grain growth during deformation. In both sets of samples, HR-EBSD analysis revealed that intragranular stress heterogeneity increased in magnitude with strain and that elevated stresses are associated with regions of high geometrically necessary dislocation density. Further analysis of the stresses and their probability distributions indicate that the stresses are imparted by long-range elastic interactions among dislocations. These characteristics indicate that dislocation interactions were the primary cause of strain hardening during transient creep. A comparison of the results to predictions from three recent models reveals that the models do not correctly predict the evolution in stress and dislocation density with strain for our experiments due to a lack of previous such data in their calibrations.

**Plain-Language Summary**

Forces from earthquakes and melting glaciers cause short-term changes to the viscosity of rocks as they flow in Earth's mantle. Recent experiments have suggested that the extent of these changes is controlled by interactions among microscopic defects, called dislocations, within the lattices of crystals that make up the rocks. To be able to predict the effects of these changes on a global scale, we need to understand how the number of, and interactions among, these defects react to sudden changes in the forces driving deformation. To this end, we deformed rocks at high temperature and pressure and characterized the evolution of these defects during a change in stress. We found that the number of dislocations increased as the stress increased, as did the strength of interactions among dislocations. Our results give more support to the idea that interactions among dislocations control the evolution of the viscosity of the rocks. These results can be used to improve models that describe short-term deformation of Earth's mantle.

# 1. Introduction

In Earth's interior, geologically rapid changes in stress occur during earthquakes and as glaciers melt. These phenomena induce periods of transient deformation in Earth's mantle in the form of postseismic creep and glacial isostatic adjustment, respectively, which can cause mantle viscosity to evolve by orders of magnitude over relatively small intervals in strain (e.g., Karato 1998; Freed et al., 2012; Simon et al., 2022; Lau et al., 2023). Such periods of transient deformation occur as the microstructure and micromechanical state of a material adjust toward a new steady state. The time and strain scales over which these transient processes occur are controlled by how quickly the key microscale state variable(s) of the deforming material can evolve in response to macroscopic changes (e.g., Kassner et al., 1986; Blum et al., 2002).

Despite the significant impact of transient processes on the short-term evolution of mantle viscosity, accurately modeling transient creep is made difficult by a lack of experimentally determined constraints on model parameters. This data gap is in part due to the limited availability of experimental data on transient creep of mantle minerals, such as olivine, for existing models to be calibrated and tested against. Only a handful of studies have focused on the evolution of viscosity with strain during transient creep of olivine, and fewer still have detailed the microstructural evolution that occurs during transient deformation. Typically, these analyses have used a Burgers rheological behavior to empirically fit the transient portion of the mechanical data from samples deformed to steady state. Post (1977) and Chopra (1997) each analyzed the mechanical response of dunite in terms of Burgers models for transient creep, while Smith and Carpenter (1987) and, more recently, Masuti and Barbot (2021) re-analyzed data from experiments on olivine single crystals and dunites using a similar framework. Additionally, these Burgers models are often connected to specific models of intergranular transient creep (Karato et al., 2021). To investigate intragranular processes, Hansen et al. (2021) performed stress-reduction tests on single crystals of olivine to characterize the nature of the transient response to changes in stress. These experiments demonstrated that stress reductions can induce a period of time-dependent reverse strain due to an internal backstress that opposes the applied stress. Hansen et al. (2021) proposed that this backstress arises from intragranular stress heterogeneity due to long-range elastic interactions among dislocations and that the evolution of this stress heterogeneity may play an important role in controlling transient creep. This stress heterogeneity has been mapped in single crystals and aggregates of olivine deformed to steady state (Hansen et al., 2021; Wallis et al., 2017, 2021). While previous studies did not characterize the evolution of stress heterogeneity, the backstress hypothesis emphasizes the need for such detailed characterization of microstructural and micromechanical evolution during transient creep. In terms of this microstructural evolution, Durham et al. (1977) and Hanson and Spetzler (1994) documented the evolution of dislocation density in natural and synthetic olivine single crystals deformed in multiple orientations and noted that dislocation density increased with strain during constant stress experiments. In contrast, Thieme et al. (2018) performed a series of similar experiments on aggregates of San Carlos olivine, but were unable to detect changes in dislocation density with strain using electron backscatter diffraction (EBSD) and transmission electron microscopy (TEM). The scarcity of microstructural data in particular makes it difficult to link models to the micromechanical processes that control the evolution of viscosity during transient creep. This connection is particularly important for developing a model that can be confidently extrapolated to mantle conditions.

To better quantify the microstructural and micromechanical evolution that occurs during transient creep in olivine aggregates, we deformed polycrystalline samples of San Carlos olivine. We characterized changes in dislocation density with increasing strain using high-angular resolution electron backscatter diffraction (HR-EBSD) and oxidation decoration. HR-EBSD exhibits lower noise levels than the conventional EBSD used by Thieme et al. (2018), and oxidation decoration characterizes larger areas than those accessible by TEM. Both these advantages increase the sensitivity to changes in dislocation density relative to this previous work. Using HR-EBSD, we also characterized the evolution of intragranular stress heterogeneity with strain. We then compare the experimental results to predictions from three recent models to assess their viability for describing the evolution of dislocation density and applied stress during transient creep. Specifically, we compare the results to the models of Holtzman et al. (2018), Mulyukova and Bercovici (2022), and Breithaupt et al. (2023). Unlike the models described previously that were empirically fit to experimental data, these models physically link microstructural changes during transient creep to the evolution in material strength, which allows us to make more direct comparisons with our results.

## 2. Methods

### 2.1 Sample Preparation

We prepared polycrystalline samples from powders of San Carlos olivine (Fo$_{90}$). Gem-quality single crystals of San Carlos olivine were crushed to produce a powder with an average particle size of 5.5 ± 2.2 µm. These powders were dried in a one-atmosphere furnace for 12 h to remove any adsorbed water at a temperature, $T$, of 1273 K and a partial pressure of oxygen of $10^{-5.2}$ Pa corresponding to that of the Ni-NiO buffer and controlled by a CO/CO$_2$ gas mixture. After drying the powders, we prepared samples for experiments by uniaxially cold pressing powders into two Ni capsules. A mass of 0.5 g of NiO powder was cold pressed into the base of each capsule to control the partial pressure of oxygen at the Ni-NiO buffer during hot pressing, and the rest of the capsule was filled with olivine powder (Meyers, 2023). Ni capsules used for cold pressing were right cylinders with an outer diameter of 14.95 mm, an inner diameter of 13.55 mm, and a height of 30 mm. We cold pressed the powders at room temperature and a pressure, $P$, of approximately 100 MPa. A Ni post with a diameter of 8 mm was included in the center of each cold press to produce thin-walled cylindrical samples. The use of thin-walled cylinders limits the radial variation in stress, and therefore microstructure, in the samples during deformation (Paterson and Olgaard, 2000; Hansen et al., 2012a). After cold pressing, samples were further densified via evacuated hot pressing at $T$ = 1523 K and $P$ = 300 MPa for approximately 2 h. During hot pressing, a vacuum was drawn on one end of the sample with a vacuum pressure of 15–30 Pa to remove trapped gasses and promote densification (Meyers, 2023).

We produced two hot presses of olivine. Both hot presses had an inner diameter of ~8 mm and an outer diameter of ~12 mm after hot pressing. One hot press resulted in a fine-grained sample, PT-1316, with grain sizes in the range 2–30 µm and a mean grain size of 4.7 µm. The other hot press resulted in a coarser-grained sample, PI-2112, with grain sizes in the range 2–155 µm and a mean grain size of 19.7 µm. Each hot press was sectioned into five cylindrical samples, approximately 3 mm in height; one sample was set aside to document the initial microstructures in each hot press, while the remaining four samples were used for deformation experiments. We assume that the initial microstructure of each deformed sample was the same as that documented from the undeformed sample from the same hot press.

## 2.2 Deformation Experiments

To prepare each deformation assembly, samples were first placed between two porous alumina spacers (Alfa Aesar, Al-25). To maintain sample alignment, the sample and surrounding spacers were placed into a Ni capsule with a length of ~17 mm, an outer diameter of 14.95 mm, and an inner diameter of 13.55 mm. The sample and spacers were stacked between alumina and zirconia pistons to center the sample in the hot zone of the furnace during deformation. Finally, the entire assembly was placed into a Fe jacket with a thickness of 0.4 mm before being inserted into the deformation apparatus.

Samples were deformed in torsion using a gas-medium deformation apparatus equipped with a torsion actuator (Paterson and Olgaard, 2000). Deformation experiments were conducted at $T = 1523$ K, $P = 300$ MPa, and a constant shear strain rate, $\dot{\gamma}$, of $1.5 \times 10^{-4}$ s$^{-1}$. The experiments were conducted at a constant strain rate to increase the duration of the initial transient in strain relative to those in constant stress experiments (Chopra 1997). Additionally, the strain rate was selected to produce as great of a peak stress as possible without the differential stress exceeding the confining pressure.

To determine the torque supported by the jackets and Ni post during transient deformation, we performed two additional experiments at the same experimental conditions listed above. One of these experiments was performed on a sample composed entirely of Fe and another on a sample composed entirely of Ni, the results of which are presented in the Supplementary Material, Figure S1. The torque measured from these two experiments was scaled based on the dimensions of the Fe jacket, Ni jacket, and Ni post from the experiments on samples of olivine. The scaled values from each of the metal components were then subtracted from the torque measured in each experiment on olivine. Typical corrections were on the order of 0.4 Nm for Fe, and 0.7 Nm and 1.4 Nm for the Ni jacket and post, respectively. At most, these values correspond to 9% of the torque measured in an experiment. These corrected torque values were then converted to shear stress, $\tau$, at the outer edge of the sample. Additionally, radial-displacement data were collected during each experiment and corrected for the compliance of the apparatus following Hansen et al. (2012). The corrected torque and radial displacement were then converted to shear stress and shear strain, $\gamma$, respectively, using the relevant equations from Paterson and Olgaard (2000). For thin-walled samples, the conversion between torque and shear stress is not sensitive to the form of transient flow law, as discussed in Supplementary Text S1. As a result, we analyzed the mechanical data using the equations presented in Paterson and Olgaard (2000) using an empirical stress exponent of $n$ = 3.1 (Wiesman et al., 2023b).

Each experiment was stopped when the target torque, and corresponding stress, was reached. We aimed to evenly space the target stresses of each experiment to best sample the variation in microstructure across the full range of available strengths. When each experiment was completed, the actuator motor was stopped, and the sample was cooled and depressurized to room temperature and pressure within 30 minutes to preserve the sample microstructure. Shear stress and shear strain were converted to equivalent stress, $\sigma$, and equivalent strain, $\varepsilon$, using the relationships of Paterson and Olgaard (2000) for a torsional deformation geometry, specifically $\sigma = \sqrt{3}\tau$ and $\varepsilon = \gamma/\sqrt{3}$, to facilitate comparison with models and samples deformed in other loading geometries.

## 2.3 Microstructural Analysis

After each experiment, the transverse section (Paterson and Olgaard, 2000, Figure 7) of each sample was polished with diamond lapping film down to a grit size of 0.5 μm followed by a chemo-mechanical polish with 40 nm colloidal silica to remove surface damage. We used three primary methods of microstructural analysis, specifically electron backscatter diffraction (EBSD), high-angular resolution electron backscatter diffraction (HR-EBSD), and dislocation decoration by oxidation, each of which is described below. The undeformed microstructure of sample PT-1316 was previously characterized by Wiesman et al. (2023a), but was re-analyzed using additional techniques in the present study.

### 2.3.1 Electron Backscatter Diffraction

Sample microstructures were mapped in a Zeiss Gemini 300 scanning electron microscope (SEM) equipped with an Oxford Instruments Symmetry EBSD detector. EBSD data were collected at an accelerating voltage of 30 kV using the AZtecHKL 4.0 data acquisition software. The microscope was calibrated for HR-EBSD following Wilkinson et al. (2006) and reference-frame conventions were validated following Britton et al. (2016). For EBSD, maps were collected with a step size of 0.5 μm over an area of 500 × 500 μm$^2$. For HR-EBSD, maps were collected with a step size of 0.25 μm over areas between 75 × 75 μm$^2$ and 150 × 150 μm$^2$. Diffraction patterns at each point in HR-EBSD maps were gathered at the maximum resolution of the detector, giving 1244 × 1024 pixels$^2$ in each pattern, and averaged over 10 frames to provide high-quality patterns. The diffraction patterns were saved for subsequent postprocessing.

Grain size and crystallographic preferred orientation (CPO) were determined from the large (500 × 500 μm$^2$) EBSD maps using the Channel5 software and MTEX toolbox for MATLAB (Bachmann et al. 2010). Grains were identified as the areas enclosed by boundaries with misorientation angles ≥ 10° between neighboring pixels and grain size was calculated as the product of the mean intercept length and a scaling factor of 1.5 (Underwood, 1970).

### 2.3.2 High-angular Resolution Electron Backscatter Diffraction

The stored diffraction patterns were analyzed using HR-EBSD, a cross-correlation based technique for determining the relative distortions of diffraction patterns within each grain (Wilkinson et al., 2006; Britton and Wilkinson, 2011, 2012). Wallis et al. (2019) provide an overview of the procedure for processing HR-EBSD data as it applies to geological materials, and their technique is summarized here.

For every map, we manually selected reference points with high pattern quality in each individual grain. 100 regions of interest (ROIs) with a size of 256 × 256 pixels were identified in each diffraction pattern. A low-pass filter was applied to each ROI in Fourier space to reduce noise (Wilkinson et al., 2006). Within each grain, every ROI was cross-correlated against the corresponding ROI at the selected reference point. This cross-correlation analysis was used to determine a field of shifts in the positions of each ROI. A displacement gradient tensor was then fit to these shifts using a robust iterative fitting procedure (Britton and Wilkinson, 2011). Following the first pass of cross-correlation, we followed the pattern remapping procedure described by Britton and Wilkinson (2012) to improve the accuracy of the elastic strain measurements for points with lattice rotations greater than 1°. The symmetric and antisymmetric components of this tensor correspond to the elastic strains and lattice rotations relative to that of the reference point, respectively (Wilkinson et al., 2006; Britton and Wilkinson, 2011, 2012).

The elastic strains determined by this method were converted into maps describing each component of the stress tensor using the elastic properties of olivine (Abramson et al., 1997). In the present study, we focus on the $\sigma_{12}$ component of the stress tensor as this component is the least modified by sectioning the sample (Wallis et al., 2019). As these stresses are relative to the unknown stress state at the reference point, the mean of $\sigma_{12}$ was calculated within each grain and subtracted from the stress at each point within that grain. In this way, this quantity represents the stress heterogeneity relative to the mean stress state of each grain (Jiang et al., 2013; Mikami et al., 2015; Wallis et al., 2017, 2019).

The spatial gradients in lattice rotations were converted into densities of geometrically necessary dislocations (GNDs). GND densities were determined by calculating the densities of each of the six dominant dislocation types in olivine required to fit the measured lattice curvature (Wallis et al., 2016). Noise in the rotation measurements result in noise in the maps of GND density and defines the noise floor that represents the minimum value of the GND density we can resolve with this technique. This noise floor is affected by the step size used in collecting the HR-EBSD maps. Because the same step size was used in collecting each map, the noise floor therefore the same across all of the HR-EBSD maps presented here. Based on the step size of 0.25 μm, we estimate a noise floor of approximately $5 \times 10^{12}$ m$^{-2}$ in the GND measurements (Wallis et al., 2016). We also note that the noise floor can vary from grain to grain based on their crystallographic orientations relative to the detector within a given map (Wallis et al., 2019).

As the stress heterogeneity and GND densities are derived from the symmetric and antisymmetric parts of the displacement gradient tensor respectively, they provide two independent measurements on the intragranular lattice distortion.

### 2.3.3 Dislocation Decoration

As an additional measurement of the dislocation density in each sample, we followed the technique outlined by Kohlstedt et al. (1976) to oxidize each sample and count the number of dislocations observed in selected areas of each sample. After HR-EBSD data were collected, a wedge was cut from each of the 10 polished samples and annealed in air for 1 h using a box furnace at $T = 900°C$. Because this temperature and oxygen partial pressure are outside of the stability field of San Carlos olivine, this technique oxidizes each sample precipitating $Fe_2O_3$, $SiO_2$, and (Mg, Fe)$SiO_3$ along dislocations near the sample surface (Kohlstedt and Vander Sande, 1975; Kohlstedt et al., 1976). After annealing, samples were re-polished on 0.5 μm diamond lapping film followed by a chemo-mechanical polish with 40 nm colloidal silica to remove the oxidation rind on the sample surface, leaving a polished surface penetrated by oxidized dislocations.

The oxidized samples were imaged using the forescattered electron (FSE) detectors mounted on the EBSD detector at 30 kV in a Zeiss Gemini 300 SEM. Dislocations appear as bright spots or streaks due to the high atomic number of the $Fe_2O_3$ precipitated along the dislocation length compared to the surrounding olivine matrix. Imaged regions in each sample were typically between 56 × 65 μm² to 65 × 75 μm² in area. For each sample, 3 to 5 images were collected and the dislocations within each image were counted manually. In each image, only free dislocations were counted, that is, we excluded dislocations that were stored in subgrain boundaries or other low-energy configurations. Free dislocations are potentially mobile or "free" to accommodate the deformation and interact with one another. Within the imaged areas we counted between 300 and 1200 dislocations per image in undeformed samples and between 13,000 and 22,000 dislocations per image in the samples deformed to the largest strain. Where possible, we imaged the same area with FSE as was analyzed with HR-EBSD, ensuring that areas imaged with both techniques experienced the same bulk stress and strain in each sample.

## 3. Results

### 3.1 Mechanical Data

Figure 1 displays the evolution of shear stress with shear strain for each experiment, while Table 1 lists the final equivalent stresses and strains reached at the end of the experiment. During each experiment, shear stress increased with strain rapidly up to $\gamma \approx 0.03$, after which the rate of increase slowed until $\gamma \approx 0.2$. For the samples that were deformed to $\gamma > 0.2$, stress remained roughly constant until the end of the experiment, indicating that the peak stress had been reached and signaling the end of the transient phase of deformation. For both the coarse-grained samples in Figure 1a and the fine-grained samples in Figure 1b, the stress-strain data for each experiment follow the same trend as the rest of the data from other samples with the same grain size. Moreover, there are no significant differences in the maximum shear stress reached or shape of the stress-strain curves between the coarse-grained and fine-grained samples.

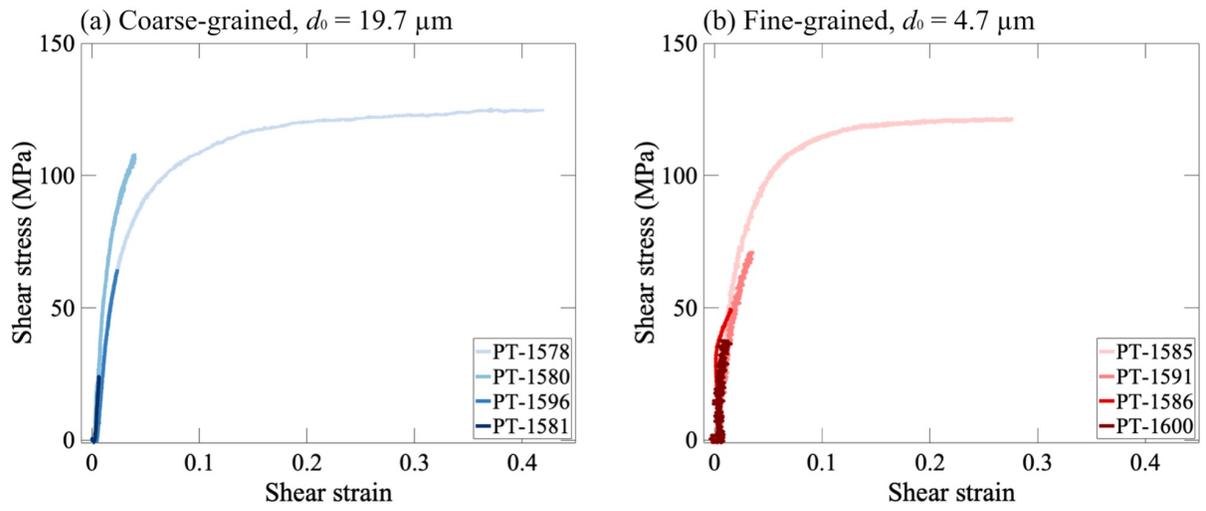

Figure 1. Shear stress versus shear strain for samples of olivine deformed in torsion for (a) coarse-grained samples from hot press PI-2112 and (b) fine-grained samples from hot press PT-1316. For both sets of samples, experiments were performed at $T = 1523$ K, $P = 300$ MPa, and $\dot{\gamma} = 1.5 \times 10^{-4}$ s$^{-1}$ or $\dot{\varepsilon} = 8.5 \times 10^{-5}$ s$^{-1}$.

| | Final equivalent strain, $\varepsilon$ | Final equivalent stress, $\sigma$ (MPa) | Mean grain size, $d$ (μm) (min.–max.)* | Dislocation density from counting, $\rho$ ($\times 10^{12}$ m$^{-2}$) | Slope of restricted second moment vs. ln($\sigma_{12}$), $m$ ($\times 10^{-2}$ GPa$^2$) |
|---|---|---|---|---|---|
| Coarse-Grained | | | | | |
| PI-2112 | 0.0 | 0 | 19.7 (2–155) | 0.16, 0.10, 0.12, 0.25, 0.13 | 1.1 |
| PT-1581 | $4.1 \times 10^{-3}$ | 41 | 21.9 (2–158) | 1.03, 0.86, 0.69, 0.67, 0.98 | 1.0 |
| PT-1596 | $1.4 \times 10^{-2}$ | 112 | 21.4 (2–177) | 1.95, 2.00, 1.94, 1.83 | 1.4 |
| PT-1580 | $2.3 \times 10^{-2}$ | 188 | 22.6 (2–161) | 2.11, 2.45, 2.58 | 1.9 |
| PT-1578 | 0.24 | 216 | 21.8 (2–166) | 2.75, 4.39, 3.86 | 5.3 |
| Fine-grained | | | | | |
| PT-1316 | 0.0 | 0 | 4.7 (2–29) | 0.08, 0.11, 0.07, 0.11 | 3.6 |
| PT-1600 | $7.4 \times 10^{-3}$ | 65 | 5.1 (2–53) | 1.05, 1.42, 1.34, 0.93 | 3.1 |
| PT-1586 | $8.7 \times 10^{-3}$ | 86 | 5.6 (2–50) | 0.54, 0.69, 0.72, 0.84, 0.87 | 3.2 |
| PT-1591 | $2.1 \times 10^{-2}$ | 123 | 6.3 (2–55) | 0.53, 0.73, 0.47, 0.58, 0.65 | 2.7 |
| PT-1585 | 0.16 | 210 | 7.2 (2–59) | 4.30, 3.63, 3.01 | 7.6 |

Table 1. Summary of mechanical and microstructural data for each sample. All samples were deformed at $T = 1523$ K, $P = 300$ MPa, and $\dot{\varepsilon} = 8.5 \times 10^{-5}$ s$^{-1}$.

* min.–max. are the minimum or maximum linear intercept, respectively, multiplied by the scaling factor of 1.5.

## 3.2 Microstructural Data

### 3.2.1 HR-EBSD

Maps of the total GND densities calculated from the HR-EBSD data are plotted in Figures 2 and 3 for the coarse-grained and fine-grained samples respectively, while the stress heterogeneities from the HR-EBSD analysis are plotted in Figures 4 and 5 for the coarse-grained and fine-grained samples respectively. The details of these results are described in the following sections.

#### 3.2.1.1 GND Density 207B

In the coarse-grained samples in Figure 2, GND density increases with strain during transient creep. The undeformed material in Figure 2a has an average GND density of $\sim 10^{13}$ m$^{-2}$, close to the estimated noise floor for our datasets. Only a handful of grains in Figure 2a have substructure in the form of linear features with a greater GND density. At $\varepsilon = 1.4 \times 10^{-2}$ in Figure 2c, linear features with large GND densities ($>10^{14}$ m$^{-2}$) are present in greater quantities than in samples deformed to smaller strains. These features correspond to subgrain boundaries with misorientation angles of $\sim 1°$. Linear features continue to increase in quantity with increasing strain, locally reaching GND densities of $>10^{15}$ m$^{-2}$. In addition, patches of high GND density ($>10^{13.5}$ m$^{-2}$) are present in grain interiors at $\varepsilon = 2.3 \times 10^{-2}$ in Figure 2d and are common throughout the mapped area at $\varepsilon = 2.4 \times 10^{-1}$ in Figure 2e.

For the fine-grained samples in Figure 3, the evolution of GND density is less obvious. There are no significant differences in GND density between samples deformed over the initial strain steps in Figures 3a–d. However, the average GND density increased in the majority of grains at the greatest strain in Figure 3e. The lack of differences between the first four strain steps is in part due to high GND densities, up to $10^{14.5}$ m$^{-2}$, present in the starting material. The large apparent GND density in the starting material is likely due to a large number of grains with orientations that produce high noise floors and therefore obscure internal GND structures (Wallis et al., 2019).

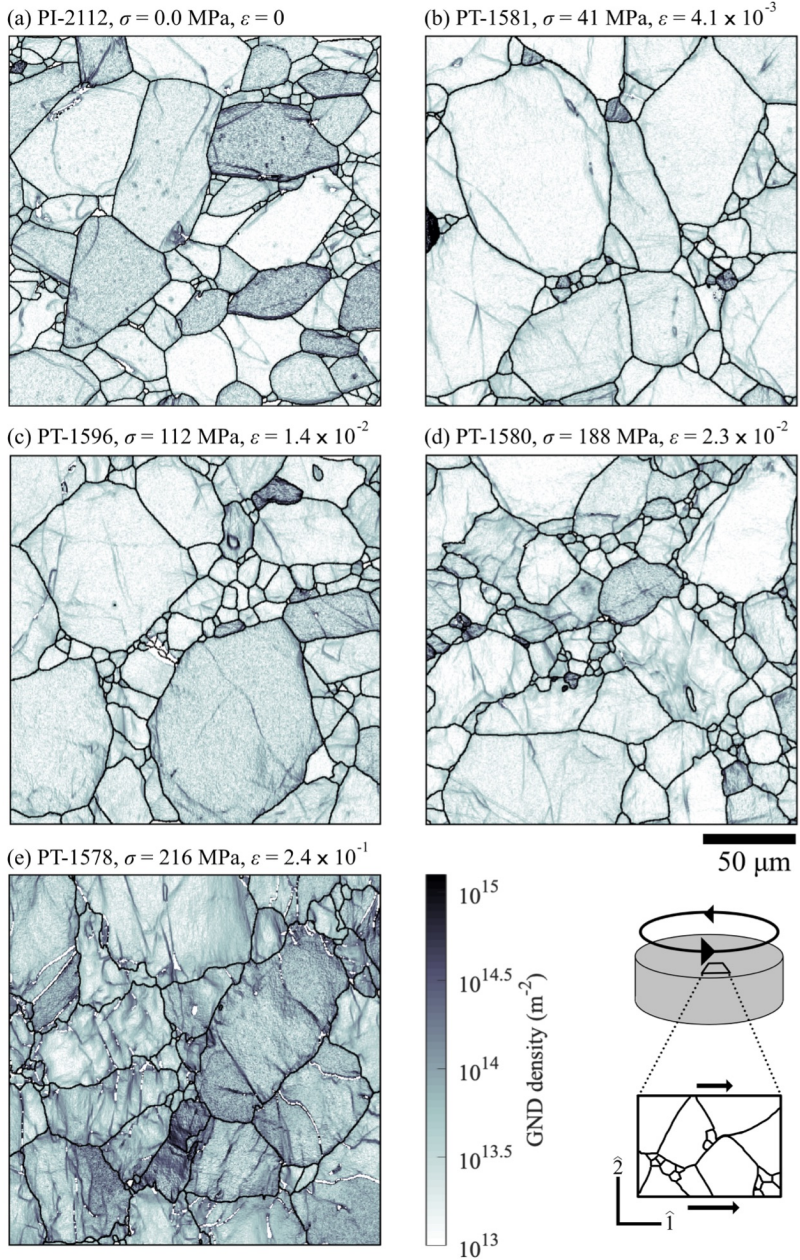

Figure 2. Total GND density calculated from HR-EBSD analysis for coarse-grained samples during transient creep in (a) the undeformed starting material, and (b–e) samples deformed to progressively larger strains. Black lines represent grain boundaries while grayscale shading represents the GND density within each grain. The orientation of the maps is displayed in the lower-right corner.

**Fine-grained samples**

(a) PT-1316, $\sigma = 0$ MPa, $\varepsilon = 0.0$
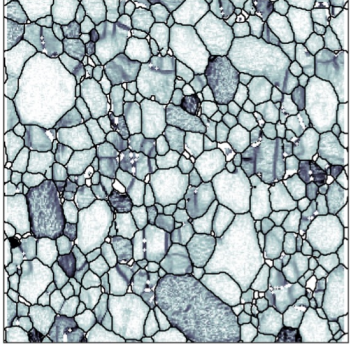

(b) PT-1600, $\sigma = 65$ MPa, $\varepsilon = 7.4 \times 10^{-3}$
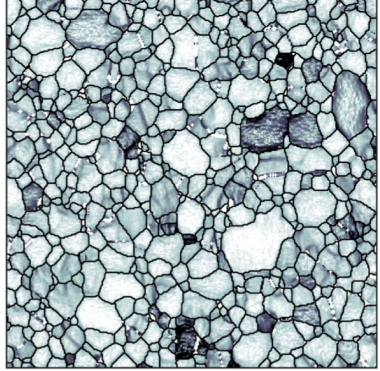

(c) PT-1586, $\sigma = 86$ MPa, $\varepsilon = 8.7 \times 10^{-3}$
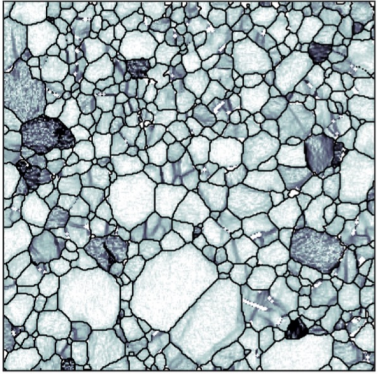

(d) PT-1591, $\sigma = 123$ MPa, $\varepsilon = 2.1 \times 10^{-2}$
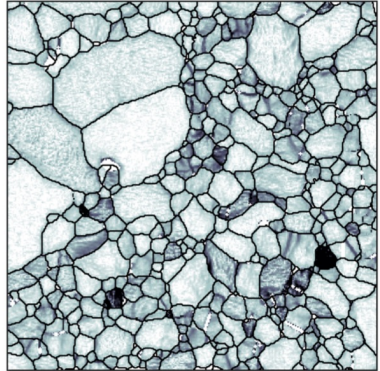

(e) PT-1585, $\sigma = 210$ MPa, $\varepsilon = 1.6 \times 10^{-1}$
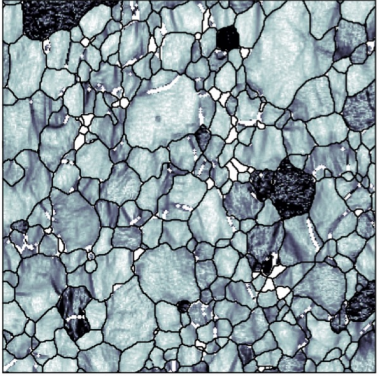

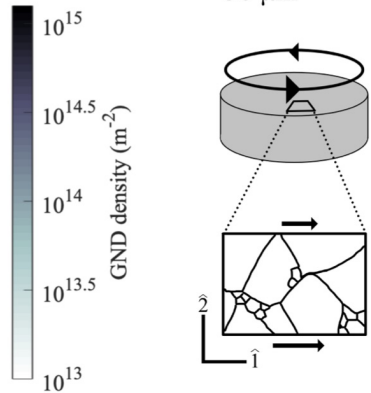

Figure 3. Total GND density calculated from HR-EBSD analysis for fine-grained samples during transient creep in (a) the undeformed starting material, and (b–e) samples deformed to progressively larger strains. Black lines represent grain boundaries while grayscale shading represents the GND density within each grain. The orientation of the maps is displayed in the lower-right corner.

### 3.2.1.2 Stress Heterogeneity

In the coarse-grained samples in Figure 4, the magnitude of intragranular stress heterogeneities increases with increasing strain during transient creep. Figure 6 demonstrates this trend by using the full width at half maximum of the intragranular stress heterogeneity distribution in each map to quantify the change in the spread of stress heterogeneities with increasing strain. There are no visible differences in intragranular stress heterogeneity between the undeformed material in Figure 4a and the sample deformed to $\varepsilon = 4.1 \times 10^{-3}$ in Figure 4b. However, patches of moderate stress heterogeneity on the order of ± 300 MPa are present in some grains by $\varepsilon = 1.4 \times 10^{-2}$ in Figure 4c. These patches are typically found near grain boundaries and extend tens of micrometers into the grain interiors. In Figure 4d, by $\varepsilon = 2.3 \times 10^{-2}$, the intensity of stress heterogeneity has increased to approximately ± 750 MPa. Patches of high stress heterogeneity occur throughout grain interiors, rather than solely near grain boundaries. In the sample deformed to the greatest strain of $\varepsilon = 2.4 \times 10^{-1}$ (Figure 4e), stress heterogeneities exceed ± 750 MPa and patches of high stress heterogeneity are found throughout grains in the mapped area. Linear features with large stress heterogeneities occur throughout the sample deformed to $\varepsilon = 2.4 \times 10^{-1}$, which likely correspond to subgrain boundaries that have developed within grains. We also note the magnitudes of the stress heterogeneities calculated with HR-EBSD are greater than both the applied stress and the confining pressure. Despite being of considerable magnitude, these stress heterogeneities are localized within grains and supported by the crystal lattice, which has a greater cohesive strength than the bulk sample (Frenkel, 1926; Hull and Bacon, 2011 pp.13).

Similar to the GND densities described above, changes in the stress heterogeneities in the fine-grained samples in Figures 5 and 6 are less obvious than in the coarse-grained samples. Moderate stress heterogeneities on the order of 400 MPa are observed near grain boundaries in the undeformed material in Figure 5a, which persist to $\varepsilon = 2.1 \times 10^{-2}$. Only at the greatest strain of $\varepsilon = 1.6 \times 10^{-1}$ in Figure 5e is an increase in the magnitude of stress heterogeneities observed. In each grain, large stress heterogeneities are present near grain boundaries. Additionally, patches of high stress heterogeneity are found in grain interiors often extending a few micrometers into each grain. Stress heterogeneities in this sample are on the order of ± 1 GPa, similar to those observed in the coarse-grained samples.

These trends in stress heterogeneity are also evident from the full width at half maximum in Figure 6c. For the coarse-grained samples, the full width at half maximum increases with increasing strain from ~0.08 GPa in the undeformed material to ~0.17 GPa in the sample deformed to the largest strain. For the fine-grained samples, the full width at half maximum remains between 0.13–0.14 GPa up to $\varepsilon = 2.1 \times 10^{-2}$ then increases to 0.23 GPa at the largest strain.

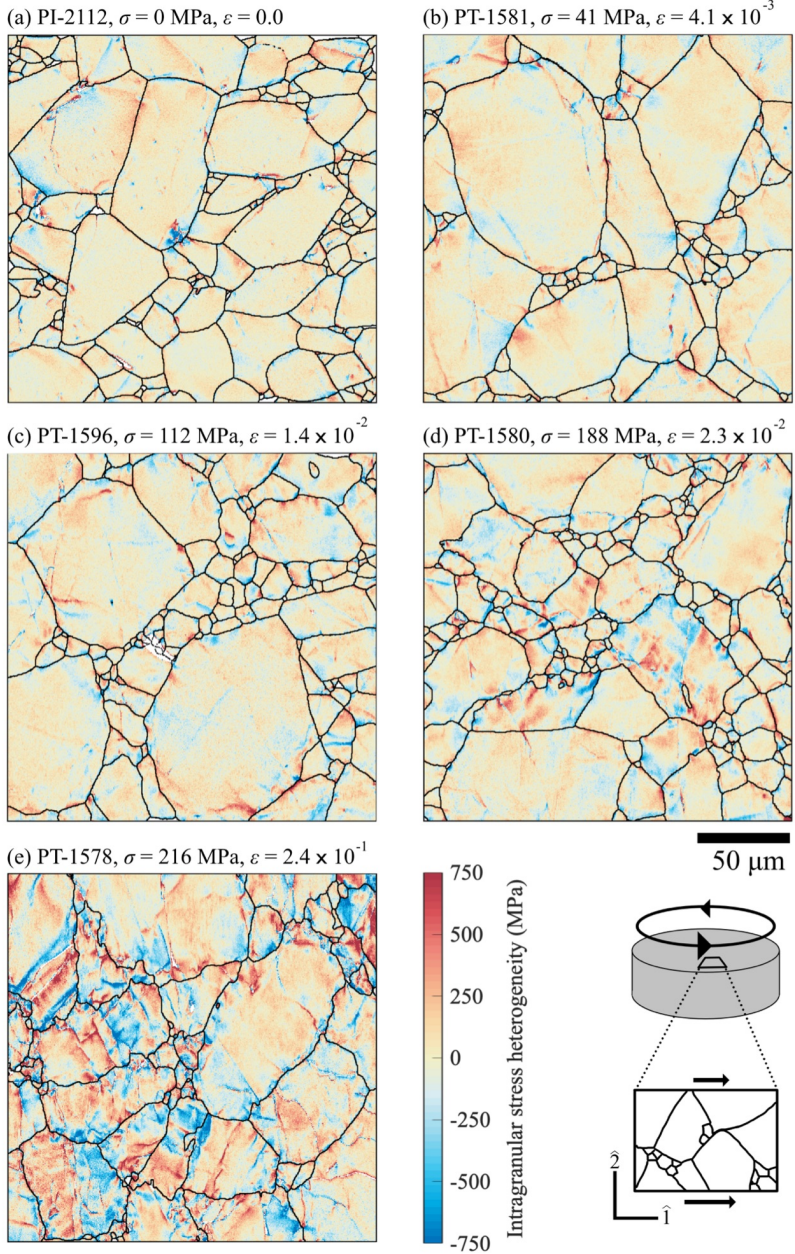

Figure 4. Intragranular stress heterogeneities calculated from HR-EBSD analysis for coarse-grained samples during transient creep. Stress heterogeneities are measured as the $\sigma_{12}$ component of the calculated stress tensor relative to the mean value of $\sigma_{12}$ in each grain. Stresses are plotted for (a) the undeformed starting material, and (b–e) samples deformed to progressively larger strains. Black lines represent grain boundaries, positive stress heterogeneities are colored in red,

and negative stress heterogeneities are colored in blue. The orientation of the maps is displayed in the lower-right corner.

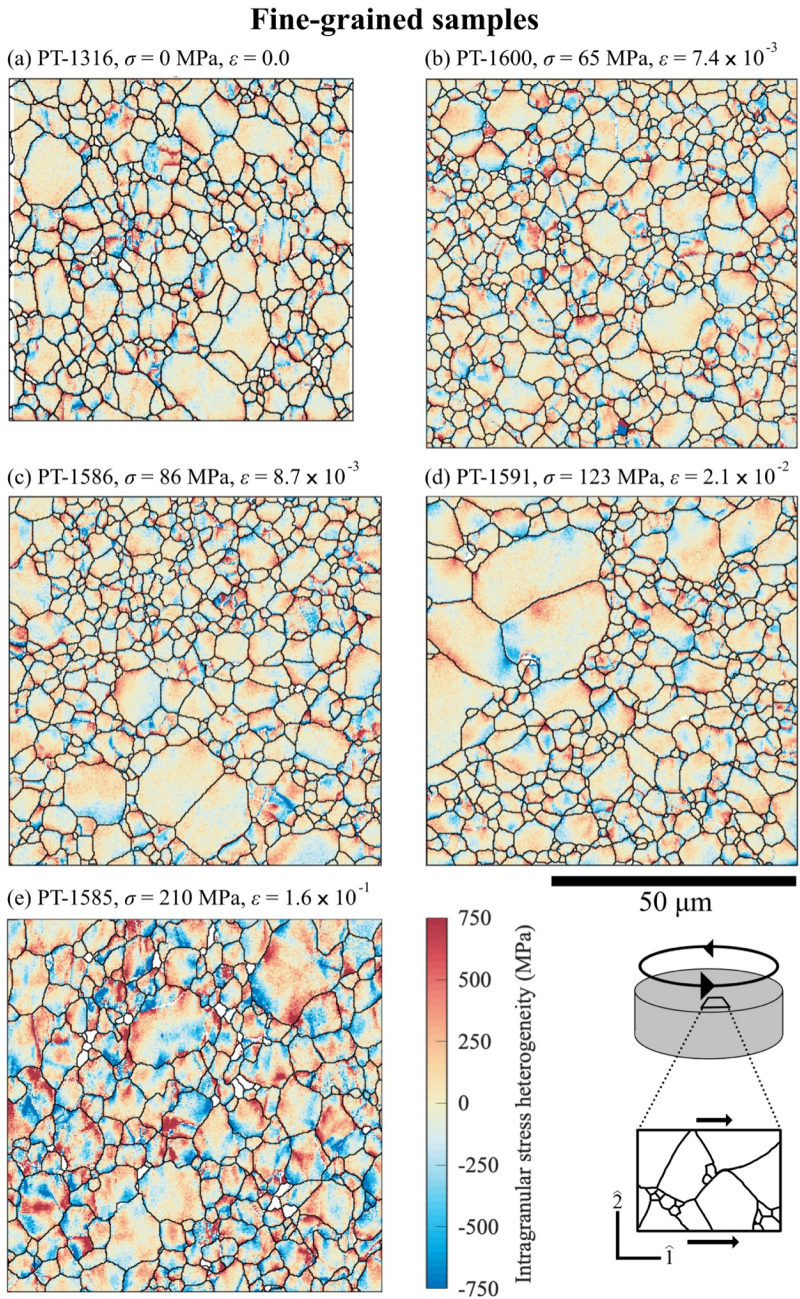

Figure 5. Intragranular stress heterogeneities calculated from HR-EBSD analysis for fine-grained samples during transient creep. Stress heterogeneities are measured as the $\sigma_{12}$ component of the calculated stress tensor relative to the mean value of $\sigma_{12}$ in each grain. Stresses are plotted for (a) the undeformed starting material, and (b–e) samples deformed to progressively larger strains.

Black lines represent grain boundaries, positive stress heterogeneities are colored in red, and negative stress heterogeneities are colored in blue. The orientation of the maps is displayed in the lower-right corner.

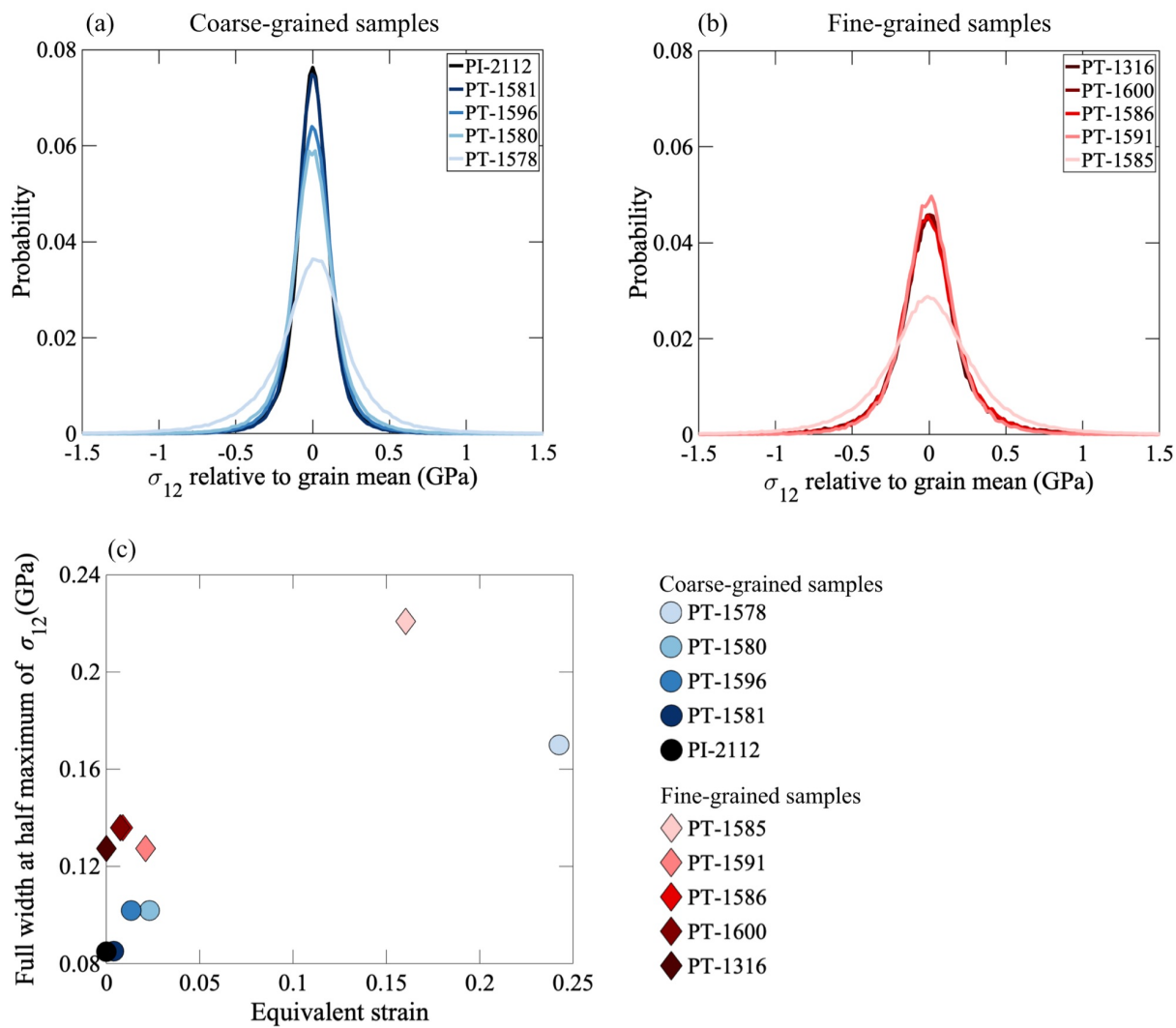

Figure 6. Evolution of the intensity of stress heterogeneity with strain. (a) and (b) Probability distributions for pixels in the maps of stress heterogeneity (Figures 4 and 5) to have a given stress value for coarse- and fine-grained samples, respectively. (c) Intensity of the stress heterogeneities are represented by the full width at half maximum of the stress-heterogeneity distribution in (a) and (b). Blue circles represent data from coarse-grained samples and red diamonds represent data from fine-grained samples.

### 3.2.2 Decorated Dislocations

Figures 7 and 8 display FSE images of the decorated coarse-grained and fine-grained samples, respectively. The dislocation densities of each sample calculated from the number of free dislocations counted in each image are listed in Table 1 and plotted against equivalent strain in Figure 9. For the coarse-grained samples, dislocation density increases with increasing strain from approximately $1.5 \times 10^{11}$ m$^{-2}$ in the undeformed material to $3.6 \times 10^{12}$ m$^{-2}$ at the largest strain and the rate of increase in dislocation density slows with increasing strain (Figures 7 and 9a). Free dislocations in these samples do not appear to be organized into slip bands and are instead dispersed throughout any given grain in a variety of orientations relative to the surface, reflecting multiple slip systems and grain orientations. This arrangement, or lack thereof, is consistent with observations of dislocations in olivine single crystals deformed at high temperatures (e.g., Wallis et al., 2017). Additionally, low-energy dislocation structures, i.e., subgrain boundaries, have formed as early as $\varepsilon = 4.1 \times 10^{-3}$ (Figure 7b).

For the fine-grained samples in Figures 8 and 9b, the evolution in dislocation density is non-monotonic. Although overall dislocation density increased from approximately $9.5 \times 10^{10}$ m$^{-2}$ in the undeformed material to $3.6 \times 10^{12}$ m$^{-2}$ in the sample deformed to the largest strain, the dislocation density measured in samples deformed to intermediate strains fluctuated between $6.0 \times 10^{11}$ m$^{-2}$ and $1.1 \times 10^{12}$ m$^{-2}$ without a clear trend. As with the coarse-grained samples above, dislocations are dispersed throughout each grain and not organized into slip bands. However, the arrangement of dislocations in the samples deformed to intermediate strains varies from grain to grain with examples of dislocations concentrated near grain boundaries, in grain centers, or spread throughout grain interiors. Specifically, clusters of dislocations are found at the center of a handful of grains, while many grains are dislocation free (Figures 8c and d). Finally, in the sample deformed to the largest strain, dislocation density has increased and dislocations are dispersed more uniformly throughout any given grain (Figure 8e).

**Coarse-grained samples**

(a) PI-2112, $\sigma = 0$ MPa, $\varepsilon = 0.0$

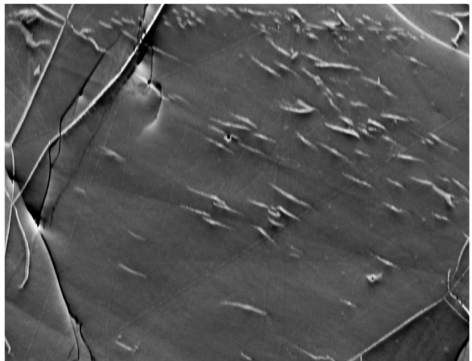

(b) PT-1581, $\sigma = 41$ MPa, $\varepsilon = 4.1 \times 10^{-3}$

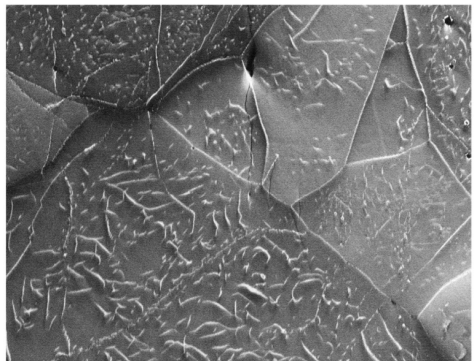

(c) PT-1596, $\sigma = 112$ MPa, $\varepsilon = 1.4 \times 10^{-2}$

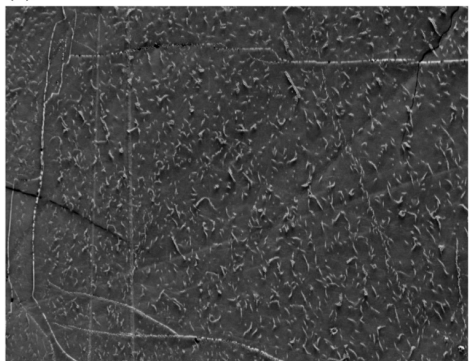

(d) PT-1580, $\sigma = 188$ MPa, $\varepsilon = 2.3 \times 10^{-2}$

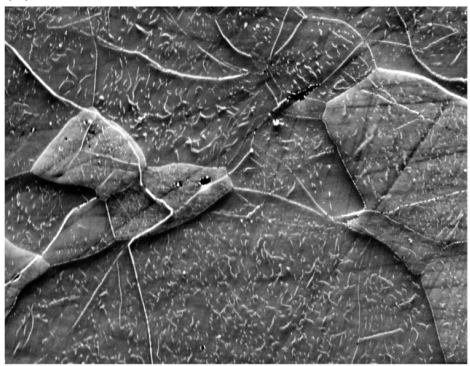

(e) PT-1578, $\sigma = 216$ MPa, $\varepsilon = 2.4 \times 10^{-1}$

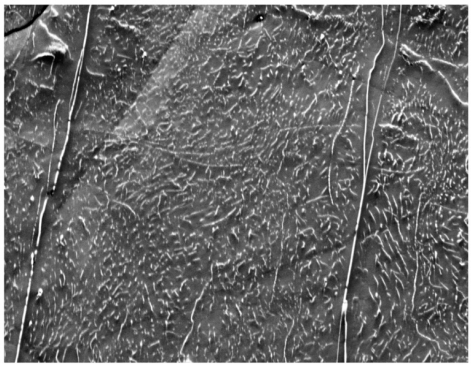

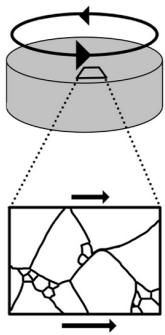

10 μm

Figure 7. FSE images of decorated dislocations in coarse-grained samples for (a) the undeformed starting material, and (b–e) samples deformed to progressively larger strains. Bright spots represent dislocations while bright line segments represent grain and subgrain boundaries. The orientation of the maps is displayed in the lower-right corner.

**Fine-grained samples**

(a) PT-1316, $\sigma = 0$ MPa, $\varepsilon = 0.0$

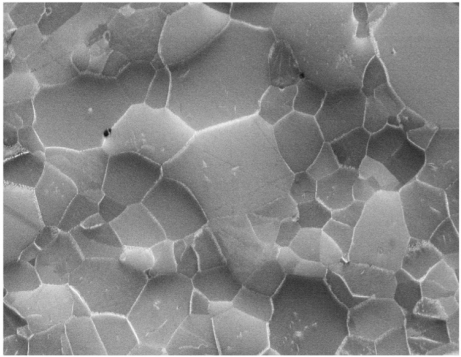

(b) PT-1600, $\sigma = 65$ MPa, $\varepsilon = 7.4 \times 10^{-3}$

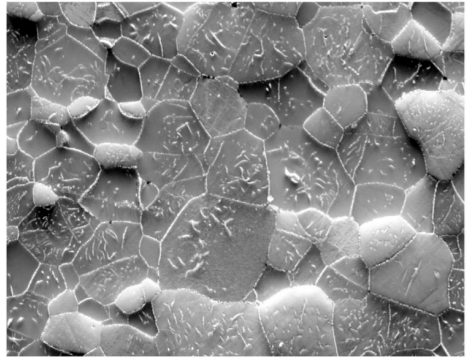

(c) PT-1586, $\sigma = 86$ MPa, $\varepsilon = 8.7 \times 10^{-3}$

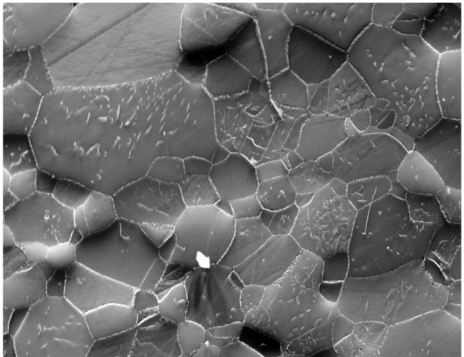

(d) PT-1591, $\sigma = 123$ MPa, $\varepsilon = 2.1 \times 10^{-2}$

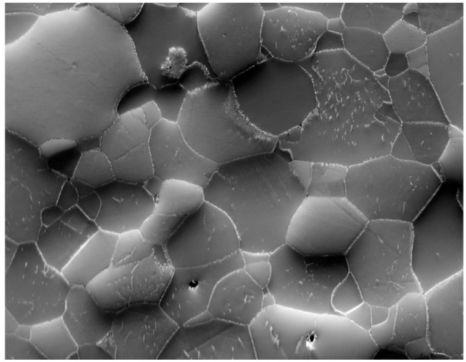

(e) PT-1585, $\sigma = 210$ MPa, $\varepsilon = 1.6 \times 10^{-1}$

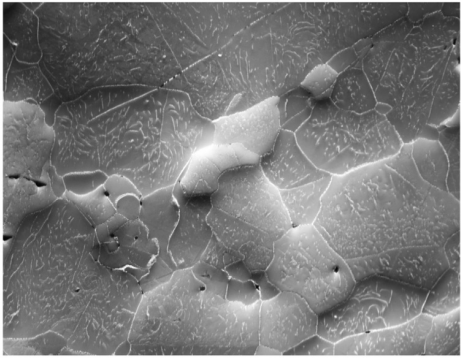

10 μm

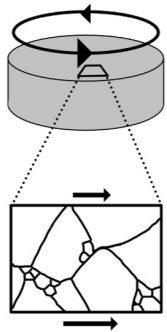

Figure 8. FSE images of decorated dislocations in fine-grained samples for (a) the undeformed starting material, and (b–e) samples deformed to progressively larger strains. Bright spots represent individual dislocations while bright line segments represent grain and subgrain boundaries. The orientation of the maps is displayed in the lower-right corner.

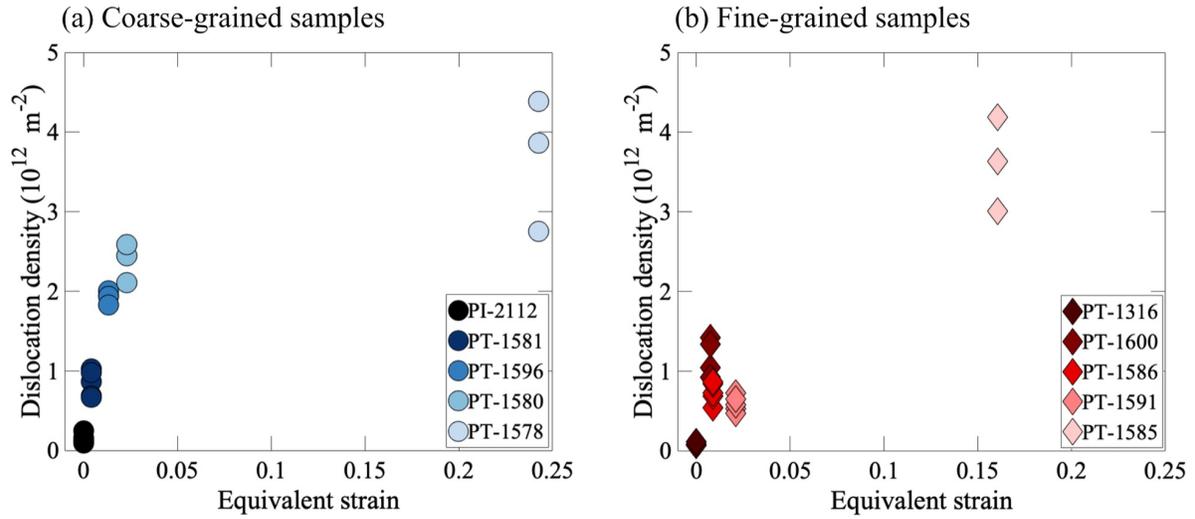

Figure 9. Dislocation density versus equivalent strain for (a) coarse-grained and (b) fine-grained samples. Dislocation density was determined by counting the free dislocations in Figures 7 and 8 and dividing by the area of each image. Multiple data points at each strain value represent dislocation densities measured from multiple images in nearby areas of the same sample.

### 3.2.3 Grain Size and CPO

The evolution of grain size with strain is plotted in Figure 10. The range of grain sizes present in the coarse-grained samples is quite large, spanning between 2 and 180 μm, however the mean grain size in each sample is between 20 and 22 μm and does not vary significantly with increasing strain. Examples of the range of grain sizes in these samples can be observed in Figures 2 and 4. In the fine-grained samples, grain size increases with strain during deformation from 4.7 μm in the undeformed material, to 7.2 μm in the sample deformed to the greatest strain. This evolution in grain size with strain is also evident in Figures 3, 5, and 8.

The evolution of CPO strength with strain, as quantified by the J-index, is plotted in Figure S2 in the supplementary material. In both the coarse- and fine-grained samples the strength of the CPO does not evolve significantly with strain, with the J-index remaining between 1 and 2 (Figure S2). These values are small compared to those for olivine deformed to shear strains greater than 3, which have undergone significant CPO development and typically have a J-index greater than 10 (Skemer et al., 2005; Hansen et al., 2012; Wiesman et al., 2023a).

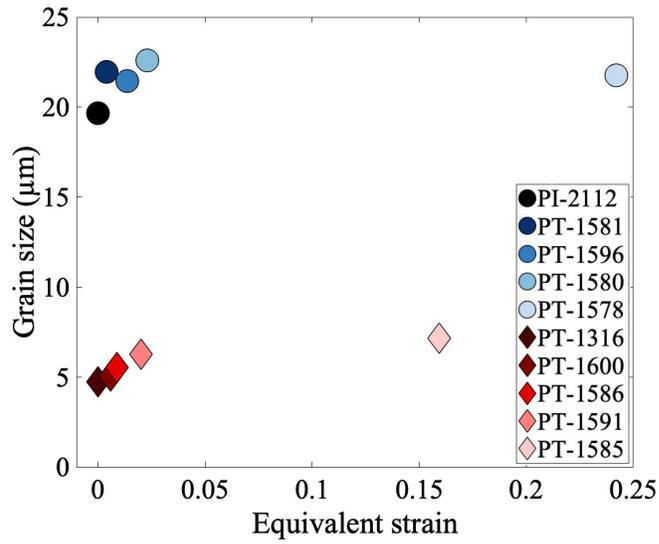

Figure 10. Evolution of grain size with strain. Coarse-grained samples are plotted as blue circles, fine-grained samples are plotted as red diamonds.

## 4. Discussion

### 4.1 Stress Evolution During Transient Creep

Multiple factors may have contributed to the stress increases that occurred during transient creep, including hardening due to dislocation interactions or grain growth if deformation is grain-size sensitive. In the following sections, we explore the role that each of these processes may have played in the strain hardening that occurred during transient creep. Although development of a CPO can affect the strength of olivine through geometrical weakening (e.g., Hansen et al., 2012), we do not discuss these effects further in the present study due to the lack of a significant change in CPO strength with strain over the strain intervals of our experiments (Figure S2).

#### 4.1.1 Stress Heterogeneity from Dislocation Stress Fields

To determine the source of the intragranular stress heterogeneity we qualitatively compare the maps of stress heterogeneity in Figures 4 and 5 to those of GND density in Figures 2 and 3, respectively. Broad patches of high GND density and localized subgrain boundaries are spatially correlated with elevated stress magnitudes, suggesting a relationship between these features.

To analyze the relationship between stress heterogeneity and dislocation density more quantitatively, we calculate the probability distributions of stress heterogeneity from the maps in Figures 4 and 5. Figures 6a and 6b present these probability distributions for the coarse- and fine-grained samples, respectively. In an analogy to the effect of stress fields around dislocations on the width and shape of X-ray diffraction spectra (Groma, 1998; Székely et al., 2001), Wilkinson et al. (2014) argued that, for the particular arrangement of straight, parallel, and non-interacting edge dislocations that are perpendicular to the sample surface, the high-stress tails of each probability distribution should decay as $P(\sigma) \propto |\sigma|^{-3}$. The normal probability plots in Figures 11a and 11b reveal that the tails of the cumulative probability distributions do indeed deviate from normal distributions at high stress magnitudes. To test for the presence of this particular relationship, we calculate the restricted second moment, $\nu_2 = \int_{-\sigma}^{+\sigma} P(\sigma')\sigma'^2 d\sigma'$, from the probability distributions. If the hypothesis that stress heterogeneity arises from stress fields around dislocations is correct, then $\nu_2$ should be proportional to $\ln(\sigma)$ at high stresses (Wilkinson et al., 2014; Kalácskska et al., 2017; Wallis et al., 2021). In Figures 11c and 11d, we plot $\nu_2$ versus $\ln(\sigma)$ for the probability distributions of $\sigma_{12}$ from each sample. The plots of $v_2$ versus $\ln(\sigma)$ have a linear segment that typically occurs between $\ln(\sigma / [1 \text{ GPa}]) = -0.6$ and $0.6$ indicating that the greatest stresses in each HR-EBSD map are due to the stress fields around dislocations. We note that, in reality, the arrangement of dislocations in each sample (Figures 7 and 8) and the interactions of stress fields among dislocations are more complicated than in the simple model used for this analysis. Nonetheless, our analysis supports the hypothesis that the stress heterogeneity is primarily due to stress fields around dislocations rather than other effects, such as anisotropic expansion and contraction in response to changes in pressure or temperature (Wallis et al. 2022). The nonlinearity of $v_2$ versus $\ln(\sigma)$ at very high stresses is potentially due to spatial averaging of the most localized and high-magnitude stresses over the interaction volume of the electron beam (Kaláckska et al., 2017), causing the measurements to deviate from the model described by Wilkinson et al. (2014). This nonlinearity may also be due to interactions among the stress fields of dislocations that are not accounted for in this simple model. The nonlinearity of $v_2$ at low stresses is also commonly observed and is related to the normally distributed portion of the probability distribution (Wilkinson et al., 2014; Wallis et al., 2021, 2022).

An additional feature of this analysis is that the slope, $m$, of the linear portion of $v_2$ versus $\ln(\sigma)$ in Figures 11b and 11d should be related to the average dislocation density via

$$m = \frac{(Gb)^2}{8\pi(1-v)^2}\rho, \quad (1)$$

where $G$ is the shear modulus, $b$ is the Burgers vector, and $v$ is Poisson's ratio (Groma, 1998; Székely et al., 2001; Wilkinson et al., 2014). Table 1 presents the values of $m$ for both the coarse- and fine-grained samples, and these values are plotted against equivalent strain in Figure 11e. For the coarse-grained samples, $m$ is approximately the same in the undeformed material and the sample deformed to $\varepsilon = 4.1 \times 10^{-3}$, but increases with increasing strain thereafter. We attribute the increase in $m$ to increasing dislocation density with strain in our samples. Using $G = 70$ GPa, $b = 5 \times 10^{-10}$ m, and $v = 0.25$ for olivine (Deer et al., 1992, pp 5; Zhao et al., 2018), we convert the measured slopes into apparent dislocation densities using Equation 1. These estimated dislocation densities range between $1.3 \times 10^{14}$ m$^{-2}$ and $6.1 \times 10^{14}$ m$^{-2}$, which are up to two orders of magnitude greater than those measured by counting the decorated dislocations (Figure 9a). This discrepancy is probably due to the restrictive nature of the assumptions about the spatial distribution of dislocations, their elastic interactions, and their orientation relative to the surface made in deriving

Equation (1). For example, the model assumes that each dislocation completely controls the stress field in its vicinity, but in reality stress fields from nearby dislocations can sum up to greater magnitudes than those produced by individual dislocations. In this scenario, fewer dislocations are required to account for the same stress magnitudes. Accounting for these effects would likely reduce the magnitude of $m$ and the associated apparent dislocation density. Additionally, dislocations counted in the decorated samples did not include dislocations stored in subgrain boundaries, whereas the analysis of the stress heterogeneity presented above is based on the full stress field, including local high stresses along subgrain boundaries with high dislocation densities. For the fine-grained samples, $m$ does not vary significantly up to $\varepsilon = 2.1 \times 10^{-2}$ and corresponds to apparent dislocation densities between $3 \times 10^{14}$ m$^{-2}$ and $4 \times 10^{14}$ m$^{-2}$. Only in the final strain step did $m$ increase, exceeding that in the coarse-grained sample deformed to the greatest strain and corresponding to an apparent dislocation density of $8.8 \times 10^{14}$ m$^{-2}$. The lack of variation in $m$ at small strains mirrors the observations in Section 3.2.2, whereby the dislocation densities measured from dislocation counting in the fine-grained samples fluctuated at small strains and were all $\lesssim 10^{12}$ m$^2$. These small fluctuations did not generate significant variation in $m$. Overall, the dislocation densities estimated from $m$ and from counting decorated dislocations display similar trends with increasing strain (Figure 11f). However, the large and variable discrepancy between dislocation densities estimated from $m$ and those from oxidation decoration suggests that $m$ provides only a qualitative indication of relative variations in dislocation density for the complex arrangements of dislocations present in deformed olivine aggregates. Nonetheless, the relationship between $v_2$ and $\ln(\sigma)$ explored here suggests that stress fields around dislocations are the primary cause of the stress heterogeneity within each sample.

Due to the link between intragranular stress heterogeneity and dislocations determined above, the distance over which stress heterogeneities extend provides information on the length scale of dislocation interactions. Visual inspection of the maps indicates that patches of high-magnitude stresses extend on the order of 10 μm within each grain (Figures 4c–e, 5e). As a more precise measure of this length scale, we analyze the autocorrelation functions of the maps of stress heterogeneity in Figures 4 and 5 to determine the typical size of patches of stress with consistent sign. The autocorrelation functions in Supplementary Figure S3 demonstrate that stress heterogeneities typically extend over a range of 4–10 μm. This distance is larger than the typical dislocation spacing of ~1 μm determined from the average dislocation densities on the order of $10^{12}$ m$^{-2}$ in each sample. Because the correlation length is larger than the average dislocation spacing, the stress heterogeneities cause long-range elastic interactions among dislocations. Increases in the intensity of long-range interactions with strain result in kinematic hardening (Weng, 1979) and therefore play a role in the observed strain hardening during deformation (Figure 1). This inference is consistent with the results of Hansen et al. (2021), who measured the backstress directly from stress-reduction tests on single crystals of olivine and determined that kinematic hardening was dominant in the hardening of their samples.

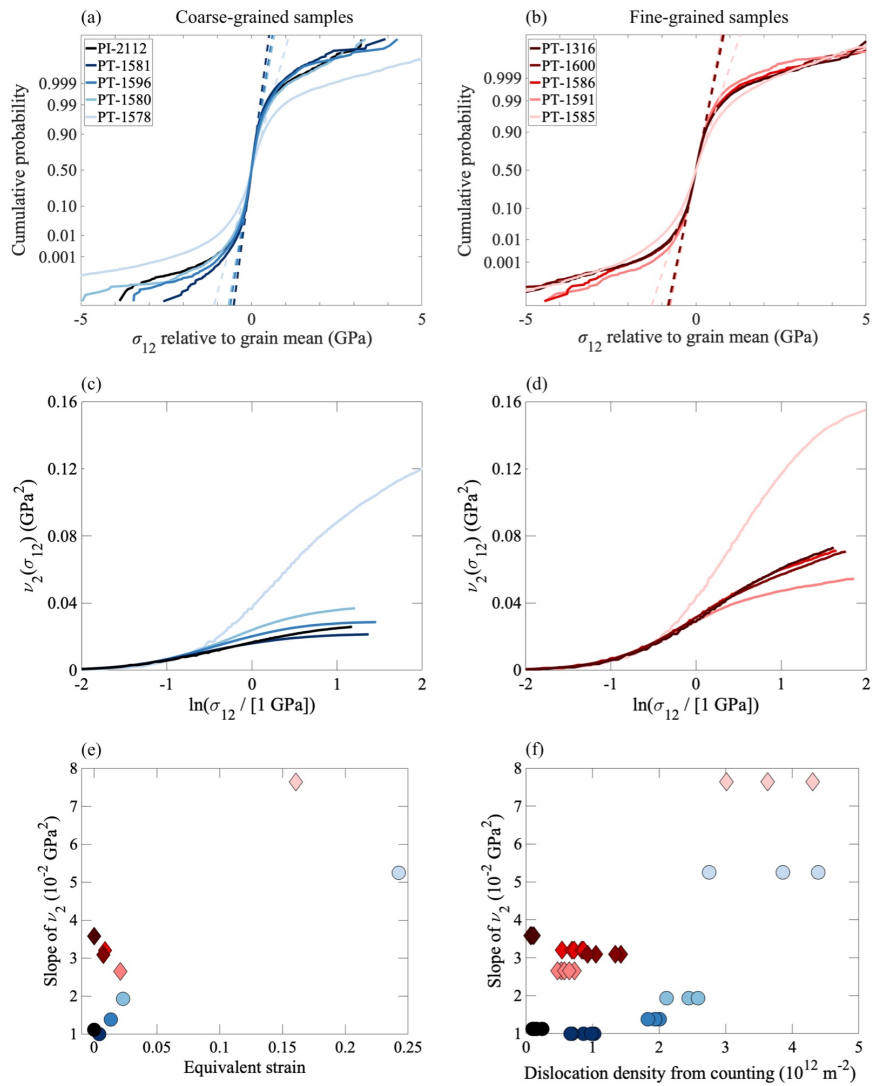

Figure 11. Analysis of the stress distributions in HR-EBSD maps. (a) and (b) Normal probability plots of the stresses relative to the mean stress in each grain for coarse- and fine-grained samples, respectively. In these plots, dashed lines represent the cumulative probability for a normal distribution of stresses, while solid lines represent the cumulative probability for the maps in Figures 4 and 5. (c) and (d) Restricted second moment calculated from the probability distributions for coarse- and fine-grained samples, respectively and plotted versus $\ln(\sigma)$. (e) The evolution of the slope of the linear portion from the plots of $v_2$ versus $\ln(\sigma)$ with strain. (f) Comparison of the dislocation densities from decorated dislocations and the slope of $v_2$ versus $\ln(\sigma)$ as a proxy for dislocation density. In panels (e) and (f) data from coarse-grained samples are plotted as blue

circles and data from fine-grained samples are plotted as red diamonds. Symbols in panels (e) and (f) are the same as those defined in Figure 9.

**4.1.2 The Effect of Grain Size on Strain Hardening and the Evolution of Dislocation Density**

As described in the previous sections, the evolution of dislocation density with strain is different between the two sets of samples. In the coarse-grained samples, dislocation density increased with strain, whereas in the fine-grained samples, dislocation density fluctuated around $1 \times 10^{12}$ m$^{-2}$ before finally increasing. Yet, as discussed in Section 3.1, both sets of samples exhibited similar strain hardening.

In the coarse-grained samples, both dislocation density and stress increased simultaneously with increasing strain. In this case, the observed strain hardening is likely a result of the increasing dislocation density. The increase in dislocation density results in increasing backstress due to long-range interactions among dislocations. As these interactions become stronger, greater stress is required to produce and move dislocations until a balance between the rate of dislocation production and recovery is reached at the peak stress.

In the fine-grained samples, dislocation density fluctuated during continuous strain hardening, yet dislocations must have been continuously produced to accommodate the strain in these samples. Using the Orowan equation, $\varepsilon = \rho b \Delta x$ (Orowan, 1940) for sample PT-1585 with an average final dislocation density of $\rho = 3 \times 10^{12}$ m$^{-2}$ and approximating the length scale, $\Delta x$, by the grain size of 7 μm would produce a strain of $\varepsilon = 0.01$, which is not enough to account for the bulk strain of $\varepsilon = 0.16$ measured in that sample. Therefore, new dislocations must have been produced throughout the transient to generate the measured bulk strain, whilst a recovery process must have removed dislocations more efficiently than in the coarse-grained samples to temporarily suppress the increase in net dislocation density. Notably, the grain size in these samples increased simultaneously over the same range in strain (Figure 10). As such, grain-boundary migration during grain growth likely provided the additional recovery mechanism that removed dislocations, counteracting the increase in dislocation density. There is some evidence for this process having occurred in Figures 8b–d; at intermediate strains, a strong contrast in dislocation density is observed across some boundaries, consistent with migrating boundaries leaving relatively pristine lattice in their wake.

Based on these observations, we can make two inferences about the role of grain size. First, the difference in mean grain size between the two sets of samples did not affect the overall evolution of dislocation density or stress. That is, both sets of samples reached approximately the same final dislocation density and the same peak stress. This observation is somewhat surprising given that the dominant deformation mechanism was likely dislocation-accommodated grain-boundary sliding (disGBS), a grain-size sensitive deformation mechanism (Hansen et al., 2011), for which we would have expected a ~90 MPa difference in final stress for the two sets of samples based on the difference in mean grain size. However, the difference in mean grain size between the two sets of samples is about the same as the scatter in the data used to calibrate the disGBS flow law (see Figure 6c of Hansen et al., 2011). Additionally, the grain-size dependance may be particularly weak for this range of grain sizes (c.f., Figure 1c of Breithaupt et al., 2023). Second, although the difference in mean grain size did not appear to affect the overall mechanical behavior, grain growth that occurred during the experiments on fine-grained samples did affect the evolution of dislocation density. As described above, dislocation density fluctuated during the initial steps in strain as migrating grain boundaries swept out dislocations. The increase in grain size alone cannot account for the total strain hardening; the disGBS flow law from Hansen et al. (2011) would only account for ~25 MPa of hardening, not the 250 MPa required to explain the strain hardening up to the largest strains explored in this study. Most of the strain hardening must have been accounted for by long-range interactions among dislocations, as in the coarse-grained samples.

**4.2 Comparison to Previous Microstructural Observations of Transient Creep**

Previous experimental studies examining microstructural evolution during transient creep of olivine are limited; Durham et al. (1977) and Hanson and Spetzler (1994) documented the evolution in dislocation density with strain during transient creep in single crystals and Thieme et al. (2018) examined this evolution in polycrystalline samples. Although experiments in each of these studies were performed under different deformation conditions and loading geometries, we compare our observations on the microstructural evolution of olivine during transient creep in the present study to these previous results to assess the overarching similarities and differences.

Hanson and Spetzler deformed single crystals of synthetic and San Carlos olivine in the $[101]_c$ and $[110]_c$ orientations under uniaxial compression at $T = 1650$ K and constant stresses each in the range 25–30 MPa. Dislocation densities revealed by oxidation decoration in their samples increased from approximately $10^{10}$ m$^{-2}$ to $10^{11}$ m$^{-2}$ over the first 1–2% strain, consistent with the strain intervals of transient creep in their experiments. Very similar results were obtained by Durham et al. (1977) on a single crystal in the $[101]_c$ orientation at $T = 1873$ K and a constant stress of 25 MPa. The initial and final dislocation densities in these studies are less than those in our samples due to the nature of the starting materials and the different experimental conditions, respectively. Nonetheless, these experiments are important in demonstrating the link between the evolution of dislocation density and transient creep in single crystals in the absence of any confounding effects due to grain boundaries (Thieme et al., 2018) and/or grain interactions (Karato, 1998, 2021) in aggregates.

Thieme et al. (2018) performed experiments on aggregates of San Carlos olivine in triaxial compression at $P = 300$ MPa and at two different temperatures of $T = 1273$ K and 1473 K. Similar to our experiments, they stopped each experiment at a different strain during the initial transient and documented the microstructures at the end of each experiment. However, they did not detect any change in dislocation density with increasing strain in their samples. Their result is in contrast to the results of Durham (1977) and Hanson and Spetzler (1994) on single crystals and results from the present study on aggregates. The likely cause of this discrepancy is the use of different analytical techniques in characterizing the microstructures. Thieme et al. (2018) used EBSD and transmission electron microscopy (TEM), whereas we used HR-EBSD and oxidation decoration. Compared to EBSD, HR-EBSD has a lower noise floor when calculating estimates of GND density (Wallis et al., 2016; 2019) allowing for features to be observed that may otherwise be obscured by noise in conventional EBSD data. Meanwhile, compared to TEM imaging, dislocation decoration allows for analysis of significantly larger areas of each sample (Kohlstedt et al., 1976) and is, therefore, more representative of the overall microstructure. The use of different techniques for analyzing microstructures in our study revealed an evolution in dislocation density that is more consistent with that observed in single crystals (Durham et al., 1977; Hanson and Spetzler, 1994).

## 4.3 Comparison to Observations at Steady State

### 4.3.1 Dislocation-density Piezometer

Similar to grain size and subgrain size, dislocation density is another metric that can be used to assess the magnitude of stress applied during deformation (Takeuchi and Argon, 1976). For olivine, Bai and Kohlstedt (1992) calibrated a dislocation-density piezometer using steady-state data from compression experiments on single crystals. Because the data from our samples were collected along the transient, for which dislocation density and stress were still evolving, the data may be expected to deviate from the relationship calibrated on steady-state data. In Figure 12, we compare our dislocation densities to the piezometer determined by Bai and Kohlstedt (1992) as well as additional dislocation densities measured at steady state by Kohlstedt and Goetze (1974), Durham et al. (1977), Hanson and Spetzler (1994), Jung and Karato (2001), Ohuchi et al. (2011), Cooper et al., (2016), and Wallis et al. (2017). The dislocation densities in our samples are broadly consistent with the steady-state piezometer. A similar result was obtained for calcite by De Bresser (1996), who observed that the dislocation density during transient creep followed the steady-state piezometric trend. Our results indicate that there is not a significant difference between the dislocation density measured during transient creep or at steady state and that dislocation density likely evolves to its piezometric value over very small strain intervals during experiments at constant strain rate.

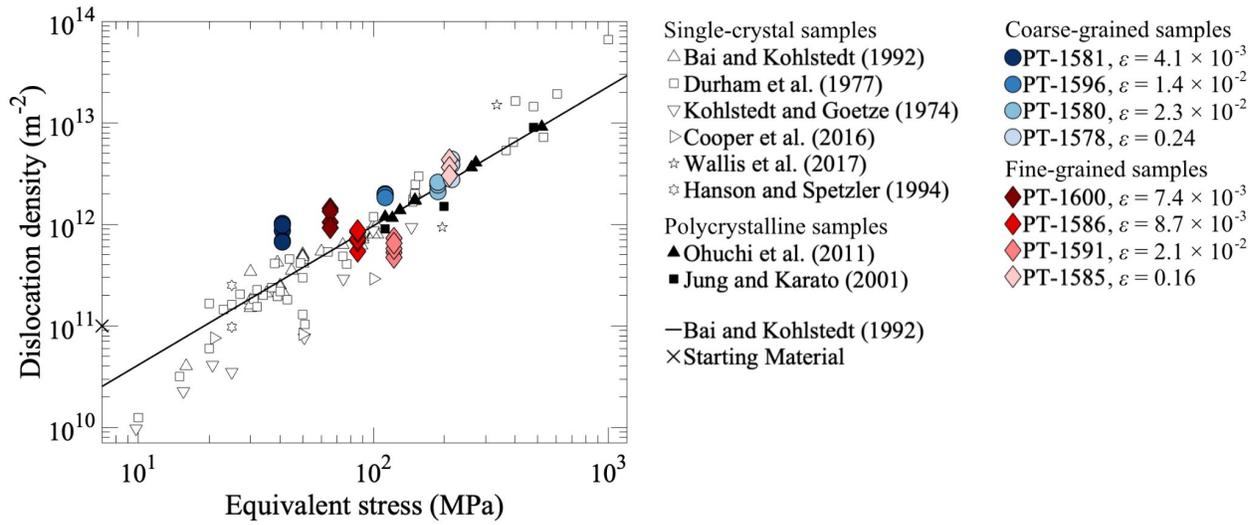

Figure 12. Dislocation density against equivalent stress. Data from experiments on single crystals and polycrystalline olivine are plotted for comparison. The dislocation-density piezometer determined by Bai and Kohlstedt (1992) is plotted as a solid black line. Data from the present study are plotted as the colored symbols with blue circles representing coarse-grained samples and red diamonds representing fine-grained samples. Dislocation density from the starting material in the present study is plotted as a black × on the left axis.

### 4.3.2 Evolution of Backstress with Applied Stress

The ratio of backstress to applied stress plays an important role in controlling the transient viscosity of a deforming material. The competition between these two quantities is evident in a recently proposed flow law of the form

$$\dot{\varepsilon}_p = A(T)\rho \sinh\left(\frac{\sigma - \sigma_\rho}{\sigma_{\text{ref}}(T)}\right), \quad (2)$$

in which $\dot{\varepsilon}_p$ is the plastic strain rate, $\sigma$ is the applied stress, $\sigma_\rho$ is the backstress due to long-range dislocation interactions, $\sigma_{\text{ref}}(T)$ represents the lattice resistance to dislocation motion, and $A(T)$ is a coefficient with an Arrhenius temperature dependence (Hansen et al., 2019; Hansen et al., 2021; Breithaupt et al., 2023). In this formulation, $\sigma_\rho$ takes the form of the Taylor stress (Taylor, 1934),

$$\sigma_\rho = \alpha G b \sqrt{\rho}, \quad (3)$$

in which $\alpha$ is a constant equal to 2.46 (Breithaupt et al., 2023). According to Equation 2, the closer the magnitude of the backstress is to that of the applied stress, the greater the change in viscosity upon a change in stress. Specialized experiments, such as stress-reduction tests, are required to directly measure backstress. Nevertheless, the magnitude of backstress can be estimated using the Taylor relationship in Equation 3 (Taylor, 1934; Thom et al., 2022; Breithaupt et al., 2023). In Figure 13, we compare the estimated backstress to the applied stress in our experiments and to previous direct measurements of backstress. It should be noted that in contrast to our estimates, previous data are measurements of the backstress at steady state. However, the close correspondence of our dislocation-density measurements from the transients to the piezometer calibrated at steady state in Figure 12 suggests that the evolution of dislocation density is an approximately quasistatic process. As backstress is linked to the dislocation density by the Taylor relationship, our estimates of backstress are likely also close to the steady-state value at any given instantaneous stress.

Our estimates of backstress follow two different ratios depending on the applied equivalent stress, with a transition occurring at an intermediate applied stress. In samples that experienced a final applied stress ≤ 120 MPa, the backstress is approximately equal to the applied stress. Whereas samples subjected to an applied equivalent stress exceeding 120 MPa instead lie close to the 1:2 line, indicating that the backstress is approximately half the applied stress. This approximate ratio of backstress to applied stress is consistent with direct measurements of backstress obtained from stress-reduction tests performed on single crystals in the $[110]_c$ and $[101]_c$ orientations (Hansen et al., 2021).

The observed transition in backstress ratio agrees with predictions made using Equation 2, which Hansen et al. (2021) hypothesized to govern both steady-state and transient creep. By coupling Equation 2 to empirical single-crystal flow laws, Hansen et al. (2021) predicted a transition between backstress ratios close to 1:1 to ratios close to 1:2 with increasing differential stresses. This transition is also predicted by Breithaupt et al. (2023) using their model of dislocation-density evolution. They predict that this transition in backstress ratio occurs due to reduced rates of dislocation recovery at low dislocation densities associated with low stresses. At lower dislocation densities, dislocations accumulate with strain until the backstress that they generate approximately balances the applied stress resulting in a backstress ratio close to unity. Whereas, at larger dislocation densities, recovery of dislocations is enhanced and dislocations are removed before the backstress fully balances the applied stress resulting in a backstress ratio less than one.

Our data support the hypothesis that deformation at differential stresses on the order of a few tens of megapascals or less has a backstress ratio close to unity. Deformation in Earth's mantle is expected to involve such low differential stresses (Ave Lallemant et al., 1980), suggesting that transient deformation in the mantle is likely to involve greater viscosity changes than those observed at higher differential stresses in the laboratory (Wallis et al., 2022).

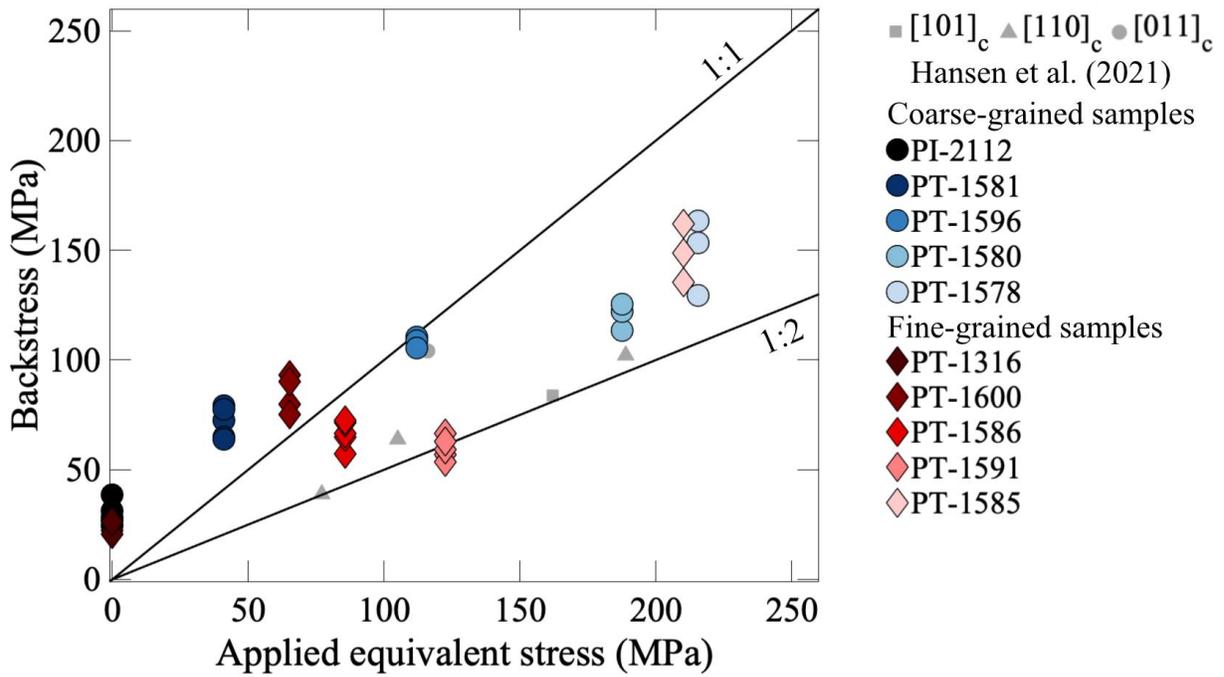

Figure 13. Backstress calculated from the Taylor equation versus the applied stress. Data from Hansen et al. (2021) for single crystals of olivine with different orientations are plotted as gray symbols and data from the present study are plotted as the colored symbols. Blue circles represent the coarse-grained samples and red diamonds represent the fine-grained samples. Backstress was predicted using the dislocation densities counted from decorated samples and the Taylor equation in Equation 3. The two black lines represent limits for the stress in which the backstress equals the applied stress (1:1) or the backstress is half the applied stress (1:2).

**4.4 Comparison to Model Predictions of Transient Creep**

Recent models have attempted to describe the co-evolution of stress and microstructural parameters, such as dislocation density and grain size, with strain. However, due to a lack of microstructural data during transient creep, these models are not necessarily well calibrated to describe the evolution of these parameters in response to changes in deformation conditions. To this end, we compare the evolution of stress and dislocation density with strain in our experiments to predictions from three recent models to identify where improvements are needed to better describe transient creep. The three models that we compare to are those of Holtzman et al. (2018), Mulyukova and Bercovici (2022), and Breithaupt et al. (2023), which are briefly summarized here. Holtzman et al. (2018) developed a thermomechanical framework resulting in a set of constitutive relations that describe the high-temperature evolution of stress, grain size, dislocation density, temperature, and strain rate. Mulyukova and Bercovici (2022) selected a number of processes that contribute to dislocation generation and recovery alongside the driving forces for changes in grain size, then composed a system of coupled equations describing the evolution of dislocation density and grain size with time in deforming rocks. Lastly, Breithaupt et al. (2023) developed their model by focusing on the mechanisms for dislocation generation and recovery as well as considering the elastic interactions among dislocations. Their efforts resulted in a pair of coupled equations: a flow law that takes the dislocation density into consideration and an equation that describes the evolution of dislocation density with time. Although empirical models based on Burgers models have been proposed to describe transient creep of dunite (e.g., Masuti and Barbot, 2021), these models only predict the mechanical evolution of stress during transient creep but do not describe the evolution of dislocation density with strain. Because the co-evolution of stress and dislocation density is an important consideration in the present study, we do not compare our results to these models. Similarly, other intergranular models for transient creep based on Burgers models predict a transition between transient and steady-state behavior over strain intervals on the order of the elastic strain (Karato et al., 2021). This prediction is not borne out by our results as transients in our samples occur over larger ranges in strain than the elastic strain. Therefore, we do not compare our results to these models either.

In Figure 14, we compare predictions from each of the three models considered to the evolution of stress and dislocation density with strain in each set of experiments. Model predictions were calculated for the experimental conditions used in the present study, specifically, $T$ = 1523 K, $P$ = 300 MPa, at a constant equivalent strain rate of $\dot{\varepsilon}$ = 8.5 × 10$^{-5}$ s$^{-1}$. We used initial conditions of $\sigma$ = 0 MPa, and the initial grain sizes and dislocation densities listed in Table 1 for samples PI-2112 and PT-1316 for the coarse- and fine-grained samples, respectively.

The model from Holtzman et al. (2018) predicts an evolution in stress that is too fast compared to the experimental results (Figure 14a), however the evolution in dislocation density predicted by their model matches the data quite well (Figure 14b). The similarity between their model predictions and the measurements of dislocation density is likely a result of their model having been calibrated to reproduce the empirical dislocation-density piezometer. Specifically, because the dislocation densities in each sample follow the piezometer during transient creep (Figure 12), their model is able to capture this evolution along the piezometer. The fast evolution in stress predicted by their model compared to the mechanical data from experiments may be due to the use of steady-state flow laws in their calculation of the stress. Additionally, their model does not consider backstress from long-range dislocation interactions, which is likely important and may provide an additional feedback to slow the evolution of stress in response to increasing dislocation density. We note that we have already accounted for the compliance of the apparatus, which can also contribute to the prolonged evolution of stress with strain.

The model of Mulyukova and Bercovici (2022) over-predicts stress by an order of magnitude (Figure 14c) and dislocation density by three orders of magnitude (Figure 14d) compared to the experimental data. This discrepancy is likely due to the low values of the average dislocation velocity used in their model. To match our experimental strain rates with these velocities requires greater dislocation densities than are observed, which in turn requires greater stresses to nucleate and move these dislocations. Their model suggests that, once the stress and dislocation density are great enough, dislocation climb becomes as fast as or faster than dislocation glide, allowing for recovery processes to more rapidly annihilate dislocations and likewise reduce the stress. However, the values of stress and dislocation density predicted by their model at steady state still significantly exceed those observed in our samples.

Finally, the model of Breithaupt et al. (2023) predicts steady-state stresses (Figure 14e) and dislocation densities (Figure 14f) close in magnitude to the experimental results. However, the predicted evolution in both stress and dislocation density during transient creep occurs too quickly over a narrow range in strain, similar to that from the model of Holtzman et al. (2018). As discussed above, we have already accounted for the compliance of the apparatus, which can contribute to the prolonged evolution of stress with strain. Instead, this rapid evolution is likely a result of the plastic modulus used in their model being poorly constrained by previous experimental data. Breithaupt et al. (2023) assumed a value of 135 GPa for the plastic modulus based on experiments conducted on single crystals and aggregates of olivine deformed at $T = 25–600°C$ (Hansen et al., 2021). However, the value of this modulus may have a temperature and/or pressure dependence that has not yet been experimentally determined. For example, values of the plastic modulus inferred from experiments on single crystals of olivine deformed at $T = 1250–1300°C$ range between 9 GPa and 99 GPa and vary with crystal orientation (Hansen et al., 2021). For the experiments in the present study, a smaller value of approximately 5 GPa would be required to match the rate of evolution in dislocation density with strain.

For each of the three models presented above, one or more important aspects of the predictions are inconsistent with the experimental data. In particular, the discrepancies between model predictions and observations are most pronounced at low strains at the onset of the transients. These discrepancies are particularly problematic because natural processes involving transient creep, such as postseismic deformation and glacial isostatic adjustment, involve small strains down to the order of $10^{-5}$ (e.g., Karato and Spetzler, 1990; Muto et al., 2019). Similar to how flow laws describing steady-state creep must typically be extrapolated in stress and strain rate to describe natural processes, models of transient creep must also be extrapolated to smaller strains than those analyzed in most laboratory experiments. As such, the accuracy of model predictions at low strains is of great importance. The discrepancies between model predictions and empirical data demand that these models be revisited to better constrain their parameters and highlight the urgent need for more experimental data on transient creep over a wider range of conditions.

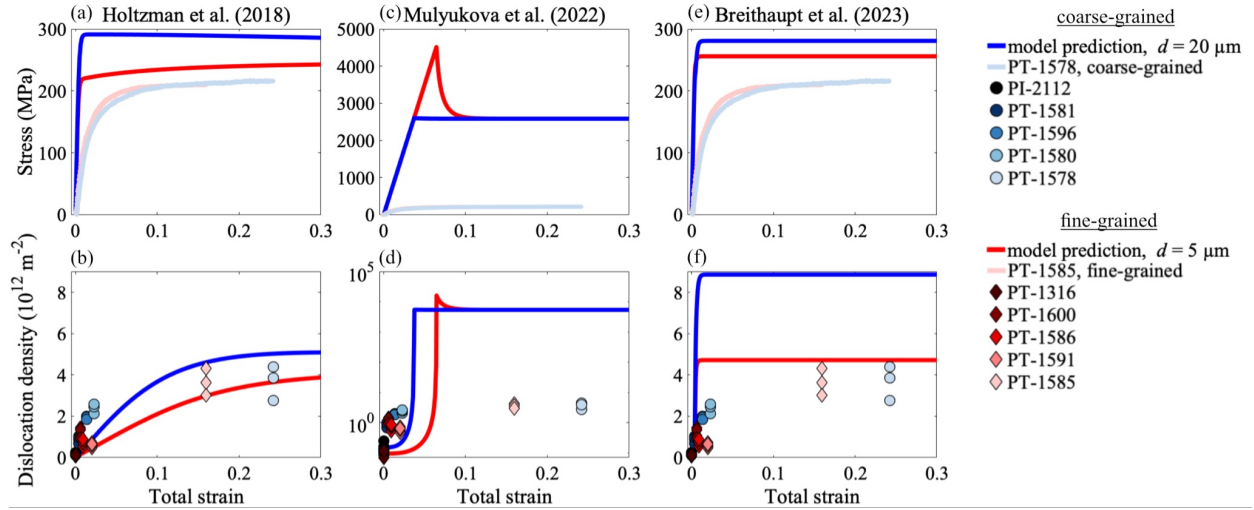

Figure 14. Model predictions for the evolution of stress and dislocation density with strain during transient creep. Predicted results from Holtzman et al. (2018) for (a) stress and (b) dislocation density, from Mulyukova and Bercovici (2022) for (c) stress and (d) dislocation density, and from Breithaupt et al. (2023) for (e) stress and (f) dislocation density. Model predictions are plotted as the blue and red curves for coarse- and fine-grained samples, respectively. Mechanical data from experiments in the present study are plotted as the light-blue and pink curves for coarse- and fine-grained samples, respectively, and dislocation densities determined for samples in the present study are plotted as blue circles and red diamonds for coarse- and fine-grained samples, respectively. For comparison with our experiments, model predictions were calculated using a constant strain rate of $\dot{\varepsilon} = 8.5 \times 10^{-5}$ s$^{-1}$ and $T = 1523$ K. For modeling of coarse-grained samples we used an initial grain size of $d = 20$ µm and initial dislocation density of $\rho = 1.5 \times 10^{11}$ m$^{-2}$ based on the values for both quantities measured in PI-2112. For modeling of fine-grained samples we used initial conditions of $d = 5$ µm and $\rho = 9.5 \times 10^{10}$ m$^{-2}$ based on the values for both quantities measured in PT-1316.

## 5. Conclusions

1) We characterized the evolution of dislocation microstructure in olivine aggregates during transient creep using oxidation decoration and HR-EBSD, both of which revealed similar trends.

    a) In coarse-grained samples ($d = 20$ μm), dislocation density increased monotonically with strain during transient creep. In our constant-strain rate experiments, measured dislocation densities matched those predicted by the dislocation-density piezometer calibrated at steady state despite being measured at various points during transient creep.

    b) In fine-grained samples ($d = 5$ μm), dislocation density did not vary significantly at the smallest strains during transient creep, but increased to approximately the same value as that in the coarse-grained samples at the largest strain. The absence of an observable change in dislocation density at small strains was likely due to simultaneous grain growth in these samples. This grain growth temporarily counteracted the increase in dislocation density but also contributed an additional potential hardening mechanism during grain-size sensitive deformation.

2) HR-EBSD maps revealed intragranular stress heterogeneity that increased in magnitude with strain during transient creep. Analysis of the restricted second moment of the stress distributions indicates that stress fields of dislocations contribute to intragranular stress heterogeneity. Additionally, stress heterogeneities in these maps typically extend over distances larger than the spacing between dislocations, demonstrating the occurrence of long-range dislocation interactions.

3) Strain hardening during transient creep was primarily due to long-range dislocation interactions. Grain growth only had a minor effect on strain hardening of the fine-grained samples during transient creep.

4) Backstress inferred from the dislocation densities equals the applied stress at stresses ≤ 120 MPa, but is about half the applied stress at stresses >120 MPa. Recent models have predicted this transition as a change in the rate-limiting processes for dislocation recovery.

5) Three recent models were unable to fully capture the evolution of stress and dislocation density in our samples during transient creep. These models need to be revisited as more data from transient creep becomes available to better match the data and make more confident extrapolations to Earth conditions.

**Data Availability**

**Acknowledgements**

The authors would like to thank Sheng Fan for helpful discussions throughout the data analysis as well as Amanda Dillman for technical assistance and the use of their facilities at the University of Minnesota. The research was supported by a UK Research and Innovation Future Leaders Fellowship grant number MR/V021788/1 awarded to DW and a grant from the Netherlands Organisation for Scientific Research, User Support Programme Space Research grant number ENW.GO.001.005 to DW, LNH, and TB. TB was also supported by a fellowship from the Royal Commission for the Exhibition of 1851. LNH was also supported by NSF grant number 2218305. Microscopy was carried out at the Wolfson Electron Microscopy Suite at the University of Cambridge, which receives funding from the Cambridge Royce facilities grant EP/P024947/1 and Sir Henry Royce Institute recurrent grant EP/R00661X/1. Experiments were carried out at the Rock and Mineral Physics Laboratory at the University of Minnesota.

**References**

Abramson, E. H., Brown, J. M., Slutsky, L. J., & Zaug, J. (1997). The elastic constants of San Carlos olivine to 17 GPa. Journal of Geophysical Research: Solid Earth, 102, 12253–12263. https://doi.org/10.1029/97jb00682

Ave Lallemant, H.G., Mercier, J-C.C., Carter, N.L., & Ross, J.V. (1980). Rheology of the upper mantle: Inferences from peridotite xenoliths. Tectonophysics, 70(1-2), 85-113. https://doi.org/10.1016/0040-1951(80)90022-0.

Bachmann, F., Hielscher, R., & Schaeben, H. (2010). Texture analysis with MTEX—Free and open source software toolbox. Solid State Phenom- ena, 160, 63–68. https://doi.org/10.4028/www.scientific.net/SSP.160.63

Bai, Q., & Kohlstedt, D. L. (1992). High-temperature creep of olivine single crystals, 2. dislocation structures. Tectonophysics, 206, 1–29. https://doi.org/10.1016/0040-1951(92)90365-d

Blum, W., Eisenlohr, P. & Breutinger, F. (2002). Understanding creep—a review. *Metall Mater Trans A* **33**, 291–303. https://doi.org/10.1007/s11661-002-0090-9

Breithaupt, T., Katz, R. F., Hansen, L. N., & Kumamoto, K. M. (2023). Dislocation theory of steady and transient creep of crystalline solids: Predictions for olivine. Proceedings of the National Academy of Sciences, 120(8), e2203448120, https://doi.org/10.1073/pnas.2203448120

Britton, T. B., Jiang, J., Guo, Y., Vilalta-Clemente, A., Wallis, D., Hansen, L. N., et al. (2016). Tutorial: Crystal orientations and EBSD — or which way is up? Materials Characterization, 117, 113–126. https://doi.org/10.1016/j.matchar.2016.04.008

Britton, T. B., & Wilkinson, A. J. (2011). Measurement of residual elastic strain and lattice rotations with high resolution electron backscatter diffraction. Ultramicroscopy, 111, 1395–1404. https://doi.org/10.1016/j.ultramic.2011.05.007

Britton, T. B., & Wilkinson, A. J. (2012). High resolution electron backscatter diffraction measurements of elastic strain variations in the presence of larger lattice rotations. Ultramicroscopy, 114, 82–95. https://doi.org/10.1016/j.ultramic.2012.01.004

Chopra, P. N. (1997). High-temperature transient creep in olivine rocks. Tectonophysics, 279, 93–111. https://doi.org/10.1016/s0040-1951(97)00134-0

Cooper, R. F., Stone, D. S., & Plookphol, T. (2016). Load relaxation of olivine single crystals. Journal of Geophysical Research: Solid Earth, 121, 7193–7210. https://doi.org/10.1002/2016jb013425

De Bresser, J. H. P. (1996). Steady state dislocation densities in experimentally deformed calcite materials: Single crystals versus polycrystals. Journal of Geophysical Research, 101, 22189–22201. https://doi.org/10.1029/96jb01759

Deer, W. A., Howie, R. A., & Zussman, J. (1992). An introduction to the rock-forming minerals (2nd ed.). Harlow, Essex, England: New York, NY: Longman Scientific & Technical.

Durham, W. B., Goetze, C., & Blake, B. (1977). Plastic flow of oriented single crystals of olivine: 2. Observations and interpretations of the dislocation structures. Journal of Geophysical Research, 82, 5755–5770. https://doi.org/10.1029/jb082i036p05755

Freed, A. M., Hirth, G., & Behn, M. D. (2012). Using short-term postseismic displacements to infer the ambient deformation conditions of the upper mantle. Journal of Geophysical Research: Solid Earth, 117, B01409. https://doi.org/10.1029/2011jb008562

Frenkel, J. (1926). Zur Theorie der Elastizitätsgrenze und der Festigkeit kristallinischer Körper. Z. Physik 37, 572–609. https://doi.org/10.1007/BF01397292

Groma, I. (1998). X-ray line broadening due to an inhomogeneous dislocation distribution. Physical Review B, 57(13), 7535–7542. https://doi.org/10.1103/physrevb.58.2969

Hansen, L. N., Zimmerman, M. E., & Kohlstedt, D. L. (2011). Grain boundary sliding in San Carlos olivine: Flow law parameters and crystallographic-preferred orientation. Journal of Geophysical Research, 116(B8), B08201. https://doi.org/10.1029/2011JB008220

Hansen, L. N., Zimmerman, M. E., & Kohlstedt, D. L. (2012). The influence of microstructure on deformation of olivine in the grain-boundary sliding regime. Journal of Geophysical Research, 117(B9). https://doi.org/10.1029/2012JB009305

Hansen, L. N., Kumamoto, K. M., Thom, C. A., Wallis, D., Durham, W. B., Goldsby, D. L., et al. (2019). Low-temperature plasticity in olivine: Grain size, strain hardening, and the strength of the lithosphere. Journal of Geophysical Research: Solid Earth, 124, 5427–5449. https://doi.org/10.1029/2018jb016736

Hansen, L., Wallis, D., Breithaupt, T., Thom, C., & Kempton, I. (2021). Dislocation creep of olivine: Backstress evolution controls transient creep at high temperatures. Journal of Geophysical Research: Solid Earth, 126, e2020JB021325. https://doi.org/10.1029/2020jb021325

Hanson, D. R., & Spetzler, H. A. (1994). Transient creep in natural and synthetic, iron-bearing olivine single crystals: Mechanical results and dislocation microstructures. Tectonophysics, 235, 293–315. https://doi.org/10.1016/0040-1951(94)90191-0

Holtzman, B., Chrysochoos, A., & Daridon, L. (2018). A thermomechanical framework for analysis of microstructural evolution: application to olivine rocks at high temperature. J. Geophys. Res. Solid Earth 123, 8474–8507.

Hull, D., & Bacon, D. J. (2011). Chapter 1 – Defects in Crystals. Introduction to Dislocations (5th ed., pp. 1-20). Butterworth-Heinemann. https://doi.org/10.1016/B978-0-08-096672-4.00001-3.

Jiang, J., Benjamin Britton, T., & Wilkinson, A. J. (2013). Mapping type III intragranular residual stress distributions in deformed copper polycrystals. Acta Materialia, 61, 5895–5904. https://doi.org/10.1016/j.actamat.2013.06.038

Jung, H., & Karato, S.-I. (2001). Water-induced fabric transitions in olivine. Science, 293, 1460-1463.

Kalácska, S., Groma, I., Borbély, A., & Ispánovity, P. D. (2017). Comparison of the dislocation density obtained by HR-EBSD and X-ray profile analysis. Applied PhysicsLetters,110, 091912.

Karato S.-I., & Spetzler, H. A. (1990). Defect microdynamics in minerals and solid-state mechanisms of seismic wave attenuation and velocity dispersion in the mantle. Reviews of Geophysics, 28(4), 337-421. https://doi.org/10.1029/RG028i004p00399

Karato, S.-I. (1998). Micro-physics of post glacial rebound. In: In P. Wu (Ed.), Dynamics of the Ice Age Earth: A modern Perspective, GeoResearch Forum (pp. 351–364). : Trans Tech Publications

Karato, S.-I. (2021). A theory of inter-granular transient dislocation creep: Implications for the geophysical studies on mantle rheology. Journal of Geophysical Research: Solid Earth, 126, e2021JB022763. https://doi. org/10.1029/2021JB022763

Kassner, M.E., Elmer, J.W. & Echer, C.J. Changes in the subboundary mesh size with creep strain in 304 stainless steel. Metall Trans A 17, 2093–2097 (1986). https://doi.org/10.1007/BF02645011

Kohlstedt, D.L., & Goetze, C. (1974). Low-stress high-temperature creep in olivine single crystals. Journal of Geophysical Research: Solid Earth, 79(14), 2045-2051. https://doi.org/10.1029/JB079i014p02045

Kohlstedt, D.L., & Vander Sande, J.B. 1975. An electron microscopy study of naturally occurring oxidation produced precipitates in iron-bearing olivines. *Contr. Mineral. and Petrol.* **53**, 13–24. https://doi.org/10.1007/BF00402451

Kohlstedt, D. L., Goetze, C., Durham, W. B., & Vander Sande, J. (1976). New technique for decorating dislocations in olivine. Science, 191, 1045–1046. https://doi.org/10.1126/science.191.4231.1045

Lau, H.C.P. (2023). Transient rheology in sea level change: implications for meltwater pulse 1A. Earth and Planetary Science Letters, 609, 118106. https://doi.org/10.1016/j.epsl.2023.118106

Mainprice, D., Barruol, G., & Ismaïl, W. (2000). The anisotropy of the Earth's mantle: From single crystal to polycrystal. In S. Karato, A. Forte, R. Liebermann, G. Masters, & L. Stixrude (Eds.), Mineral physics and seismic tomography: From atomic to global, AGU geophysical monograph (Vol. 117, pp. 237–264).

Masuti, S., & Barbot, S. (2021). MCMC inversion of the transient and steady-state creep flow law parameters of dunite under dry and wet conditions. Earth Planets and Space, 73(1), 1–21. https://doi.org/10.1186/s40623-021-01543-9

Meyers, Cameron. (2023). Experimental Deformation of Olivine Aggregates. Retrieved from the University of Minnesota Digital Conservancy, https://hdl.handle.net/11299/260654.

Mikami, Y., Oda, K., Kamaya, M., & Mochizuki, M. (2015). Effect of reference point selection on microscopic stress measurement using EBSD. Materials Science and Engineering A, 647, 256–264. https://doi.org/10.1016/j.msea.2015.09.004

Mulyukova, E., & Bercovici, D. (2022). On the co-evolution of dislocations and grains in deforming rocks. Physics of Earth and Planetary Interiors, 328, 106874, https://doi.org/10.1016/j.pepi.2022.106874

Muto, J., Moore, J. D. P., Barbot, S., Iinuma, T., Ohta, Y., & Iwamori, H. (2019). Coupled afterslip and transient mantle flow after the 2011 Tohoku earthquake. Science Advances, 5, eaaw1164. https://doi.org/10.1126/sciadv.aaw1164

Ohuchi, T., Kawazoe, T., Nishihara, Y., Nishiyama, N., & Irifune, T. (2011). High pressure and temperature fabric transitions in olivine and variations in upper mantle seismic anisotropy. Earth and Planetary Science Letters, 304 (1–2), 55-63. https://doi.org/10.1016/j.epsl.2011.01.015

Orowan, E. (1940). Problems of plastic gliding. Proceedings of the Physical Society, 52(1), 8. 10.1088/0959-5309/52/1/303

Paterson, M., & Olgaard, D. (2000). Rock deformation tests to large shear strains in torsion. Journal of Structural Geology, 22(9), 1341–1358. https://doi.org/10.1016/S0191-8141(00)00042-0

Post, R.L., Jr. (1977). High-temperature creep of Mt. Burnett dunite. Tectonophysics, 42, 75-110. https://doi.org/10.1016/0040-1951(77)90162-7

Simon, K. M., Riva, R. E. M., & Broerse, T. 2022. Identifying geographical patterns of transient deformation in the geological sea level record. Journal of Geophysical Research: Solid Earth, 127, e2021JB023693. https://doi.org/10.1029/2021JB023693

Skemer, P., Katayama, I., Jiang, Z., & Karato, S. (2005). The misorientation index: Development of a new method for calculating the strength of lattice-preferred orientation. Tectonophysics, 411(1–4), 157–167. https://doi.org/10.1016/j.tecto.2005.08.023

Smith, B.K., & Carpenter, F.O. (1987). Transient creep in orthosilicates. Physics of the Earth and Planetary Interiors, 49, 314-324. https://doi.org/10.1016/0031-9201(87)90033-1

Székley, F., Groma, I., & Lendvai, J. (2001). Statistic properties of dislocation structures investigated by X-ray diffraction. Materials Science and Engineering: A, 309–310, 352-355. https://doi.org/10.1016/S0921-5093(00)01629-4

Takeuchi, S., & Argon, A. S. (1976). Steady-state creep of single-phase crystalline matter at high temperature. Journal of Materials Science, 11(8), 1542–1566. https://doi.org/10.1007/BF00540888

Taylor, G. I. (1934). The Mechanism of Plastic Deformation of Crystals. Part I. Theoretical. Proceedings of the Royal Society of London. Series A, Containing Papers of a Mathematical and Physical Character, 145, 362–387.

Thieme, M., Demouchy, S., Mainprice, D., Barou, F., & Cordier, P. (2018). Stress evolution and associated microstructure during transient creep of olivine at 1000–1200 c. Physics of Earth and Planetary Interiors 278, 34–46.

Thom, C. A., Hansen, L. N., Breithaupt, T., Goldsby D. L., & Kumamoto, K. M. (2022). Backstresses in geologic materials quantified by nanoindentation load-drop experiments. Philosophical Magazine, 102, 1974-1988. https://doi.org/10.1080/14786435.2022.2100937

Underwood, E. (1970), Quantitative Stereology, Addison-Wesley, Reading, Mass.

Wallis, D., Hansen, L. N., Britton, T. B., & Wilkinson, A. J. (2016). Geometrically necessary dislocation densities in olivine obtained using high-angular resolution electron backscatter diffraction. Ultramicroscopy, 168, 34–45. https://doi.org/10.1016/j.ultramic.2016.06.002

Wallis, D., Hansen, L. N., Britton, T. B., & Wilkinson, A. J. (2017). Dislocation interactions in olivine revealed by HR-EBSD. Journal of Geophysical Research: Solid Earth, 122, 7659–7678. https://doi.org/10.1002/2017jb014513

Wallis, D., Hansen, L. N., Britton, T. B., & Wilkinson, A. J. (2019). High-angular resolution electron backscatter diffraction as a new tool for mapping lattice distortion in geological minerals. Journal of Geophysical Research: Solid Earth, 124, 6337–6358. https://doi.org/10.1029/2019JB017867

Wallis, D., Hansen, L., Wilkinson, A. J., & Lebensohn, R. A. (2021). Dislocation interactions in olivine control postseismic creep of the upper mantle. Nature Communications, 12, 3496. https://doi.org/10.1038/s41467-021-23633-8

Wallis, D., Sep, M., & Hansen, L.N. (2022). Transient creep in subduction zones by long-range dislocation interactions in olivine. Journal of Geophysical Research: Solid Earth, 127, e2021JB022618. https://doi.org/10.1029/2021JB022618

Weng, G.J. (1979). Kinematic Hardening Rule in Single Crystals. International Journal of Solids and Structures, 15(11), 861-870. https://doi.org/10.1016/0020-7683(79)90055-6.

Wiesman, H. S., Zimmerman, M. E. ,& Kohlstedt, D. L. (2023a). The effect of secondary-phase fraction on the deformation of olivine + ferropericlase aggregates: 1. Microstructural evolution. Journal of Geophysical Research: Solid Earth, 128, e2022JB025723. https://doi.org/10.1029/2022JB025723

Wiesman, H. S., Zimmerman, M. E., & Kohlstedt, D. L. (2023b). The effect of secondary-phase fraction on the deformation of olivine + ferropericlase aggregates: 2. Mechanical behavior. Journal of Geophysical Research: Solid Earth, 128, e2022JB025724. https://doi.org/10.1029/2022JB025724

Wilkinson, A. J., Meaden, G., & Dingley, D. J. (2006). High-resolution elastic strain measurement from electron backscatter diffraction patterns: New levels of sensitivity. Ultramicroscopy, 106(4-5), 307–313. https://doi.org/10.1016/j.ultramic.2005.10.001

Wilkinson, A. J., & Britton, T. B. (2012). Strains, planes, and EBSD in materials science. Materials Today, 15(9), 366–376. https://doi.org/10.1016/s1369-7021(12)70163-3

Zhao, Y.-H., Zimmerman, M. E., & Kohlstedt, D. L. (2018). Effect of iron content on the creep behavior of olivine: 2. Hydrous conditions. Physics of the Earth and Planetary Interiors, 278, 26–33. https://doi.org/10.1016/j.pepi.2017.12.002

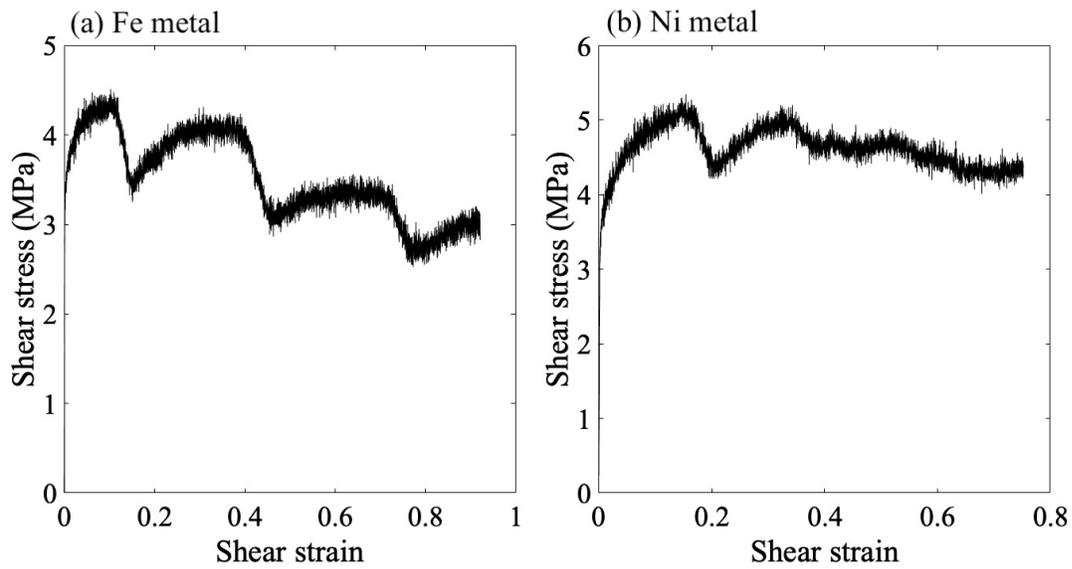

**Figure S1.** Shear stress versus shear strain for (a) Fe and (b) Ni metal. As described in Section 2.2 of the main text, the mechanical data for these two metals was used to correct the strength of our olivine samples for the strength of the jacketing materials used in each experiment. Torque data for both Fe and Ni were scaled based on sample dimensions to apply corrections to each sample. This data was used instead of a flow law to account for the transient evolution in strength of the jacketing materials at small strains when applying jacket corrections.

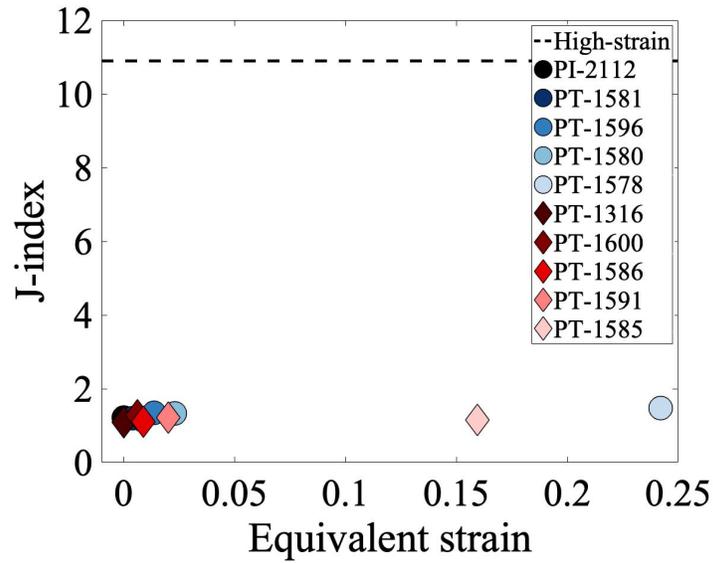

**Figure S2.** Evolution of crystallographic preferred orientation (CPO) strength with strain during transient creep. The strength of olivine CPOs is characterized using the J-index, which ranges from 1 for a random distribution to $\infty$ for a single-crystal (Mainprice et al., 2000). The J-index for Coarse-grained samples are plotted as blue circles and for fine-grained samples as red diamonds. The J-index for a high-strain olivine sample with a strong CPO (Wiesman et al., 2023a) is plotted as a dashed line for comparison.

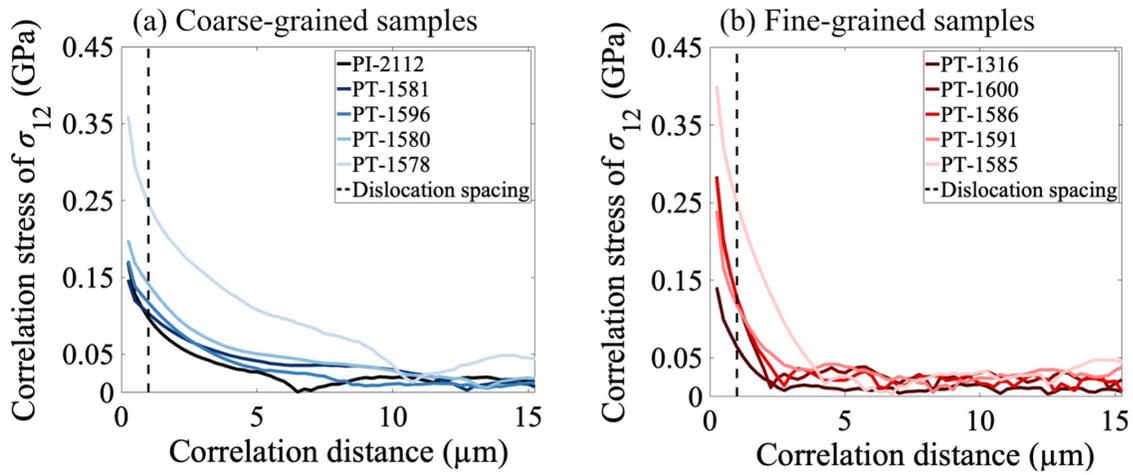

**Figure S3.** Autocorrelation function of $\sigma_{12}$ versus the correlation distance for (a) coarse-grained samples and (b) fine-grained samples. The autocorrelation function was calculated from the maps of stress heterogeneity in Figures 4 and 5. The curves plotted in (a) and (b) represent vertical traces from the center of each autocorrelation map. The vertical dashed line represents the dislocation spacing calculated from the average dislocation density in the samples deformed to the highest strain.

**Text S1: Uncertainty in the calculation of stress during transient creep**

For experiments performed in a torsion geometry, conversion between torque and shear stress typically follows the analysis outlined in Paterson and Olgaard (2000), which assumes that the sample is at mechanical steady-state. However, the mechanical data in the present study examines transient creep, which motivates revisiting this conversion. The conversion is fundamentally sensitive to the radial distribution of shear stress, which may be modified by the elastic-viscous transition and a transition in the stress-sensitivity of the viscous rheology.

From Equation 11 in Paterson and Olgaard (2000), the ratio $\kappa$ between the torque applied to the sample at mechanical steady-state, $M$, to the shear stress, $\tau$, at the outermost edge of the sample is given by

$$\kappa = \frac{M}{\tau} = \frac{\pi}{4(3+\frac{1}{n})} \frac{d_o^{3+\frac{1}{n}} - d_i^{3+\frac{1}{n}}}{d_o^{\frac{1}{n}}}, \tag{S1}$$

in which, $d_i$ and $d_o$ are the inner and outer diameters of the sample respectively and $n$ is a parameter that encapsulates the radial distribution of shear stress. Equation 11 is derived by assuming that at a given radius $r$ in the sample, the shear stress varies in proportion to $r^{1/n}$. For $n = 1$, the shear stress increases linearly with radius, but in the limit $n \to \infty$, shear stress is constant over the sample radius. For a sample with a power-law rheology at mechanical steady-state, $n$ can be identified as the stress exponent. However, we note that the application of Equation S1 is not limited to viscous rheologies. For example, in an elastic sample, the shear stress increases linearly with radius and/or strain, and thus the conversion factor may be calculated from Equation S1 with $n = 1$.

In our experiments, the samples are first loaded elastically before transitioning to a viscous rheology. For olivine at steady state, this viscous rheology has a stress exponent of 3.1 (Wiesman et al., 2023b). Thus, the radial distribution of shear stress must transition from that governed by elasticity to that governed by the steady-state. In addition, the stress-sensitivity of the viscous rheology may evolve with strain during transient creep. Fits to experimental data (Hansen et al., 2021; Masuti and Barbot, 2021), and microphysical theory (Breithaupt et al., 2023) suggest that lower stress exponents apply during transient creep, potentially as low as $n = 1$. Based on these observations regarding the evolution of $n$, we infer that the radial distribution of shear stresses likely transitions from that reflecting $n = 1$ as the sample is elastically loaded and during incipient viscous deformation, to $n = 3.1$ at mechanical steady-state.

To assess the response of our calculated stresses to the evolution of stress sensitivity, we examine the ratio between the conversion factors in Equation S1 for a power-law rheology, with a shear stress distribution governed by $n > 1$, and linear distribution of shear stresses with $n = 1$

$$\frac{\kappa_{\text{power law}}}{\kappa_{\text{linear}}} = \left(\frac{4}{3 + \frac{1}{n}}\right) \left(\frac{d_o}{d_o^{1/n}}\right) \left(\frac{d_o^{3+1/n} - d_i^{3+1/n}}{d_o^4 - d_i^4}\right). \tag{S2}$$

To clarify the role of the radial thickness of the sample, we define the parameter $x$ as the fractional width of the sample relative to the outer radius, i.e., $d_i = d_o(1 - x)$. For the typical sample dimensions used in the present study, $x \approx 1/3$. Substituting this definition of $x$ into Equation S2 and expanding in powers of $x$, we find

$$\frac{\kappa_{\text{power law}}}{\kappa_{\text{linear}}} = 1 + \frac{n-1}{2n}x + \mathcal{O}(x^2). \tag{S3}$$

This approximate relationship illustrates that the sensitivity of the conversion factor to $n$ is weak, and that in the limit of an infinitesimally thin-walled sample ($x \to 0$), the conversion factor is independent of any changes to stress-sensitivity.

Substituting our sample dimensions and the known empirical steady-state stress exponent into Equation S2, we find that the ratio $\frac{\kappa_{\text{power law}}}{\kappa_{\text{linear}}}$ is 1.11. This value indicates that the transient evolution of the stress-sensitivity could lead to an 11% uncertainty in inferred stresses, an uncertainty on the order of the jacket corrections to the mechanical data. In our discussion, we also consider the slope of the stress-strain curve; our calculated slopes also have the same uncertainty.